\documentclass[conference]{IEEEtran}
\IEEEoverridecommandlockouts
\usepackage{cite}
\usepackage{amsmath,amssymb,amsfonts}
\usepackage{algorithmic}
\usepackage{graphicx}
\usepackage{textcomp}
\usepackage{xcolor}

\usepackage{subfig}
\usepackage[ruled,vlined,linesnumbered, noend]{algorithm2e}
\graphicspath{ {figures/} }
\usepackage[export]{adjustbox}
\usepackage{multirow}
\usepackage{mathtools}

\DeclarePairedDelimiter\floor{\lfloor}{\rfloor}

\def\BibTeX{{\rm B\kern-.05em{\sc i\kern-.025em b}\kern-.08em
    T\kern-.1667em\lower.7ex\hbox{E}\kern-.125emX}}

\begin{document}

\title{FastVA: Deep Learning Video Analytics Through Edge Processing and NPU in Mobile}

\author{
\IEEEauthorblockN{Tianxiang Tan and Guohong Cao}
\IEEEauthorblockA{
	Department of Computer Science and Engineering \\
	The Pennsylvania State University\\
	Email: \{txt51, gxc27\}@psu.edu}
}

\maketitle

\begin{abstract}
	Many mobile applications have been developed to apply deep learning for video analytics. Although these advanced deep learning models can provide us with better results, they also suffer from the high computational overhead which means longer delay and more energy consumption when running on mobile devices.
	To address this issue, we propose a framework called FastVA, which supports deep learning video analytics through edge processing and Neural 
	Processing Unit (NPU) in mobile.  The major challenge is to determine when to offload the computation and when to use NPU. 
	Based on the processing time and accuracy requirement of the mobile application, we study two problems: 
	\textit{Max-Accuracy} where the goal is to maximize the accuracy under some time constraints, and 
	\textit{Max-Utility} where the goal is to maximize the utility which is a weighted function of processing time and accuracy. 
	We formulate them as integer programming problems and propose heuristics based solutions. 
	We have implemented FastVA on smartphones and demonstrated its effectiveness through extensive evaluations.
	
\end{abstract}

\section{Introduction}
Deep learning techniques, such as Convolutional Neural Network (CNN), have been successfully applied to various computer vision and natural language processing 
problems, and some of the advanced models can even outperform human beings in some specific datasets \cite{he-iccv15}.
Over the past few years, many mobile applications have been developed to apply CNN models for video analytics.
For example, Samsung Bixby can help users extract texts showing up in the video frames.
Although these advanced CNN models can provide us with better results, they also suffer from the high computational overhead 
which means long delay and more energy consumption when running on mobile devices.

Most existing techniques \cite{chen-sensys15, chen-sensys18, teerapittayanon-icdcs17, ran-infocom18, han-mobisys16} address this problem through computation offloading. By offloading data/video to the edge server and letting the edge server run these deep learning models, energy and processing time can be saved. However, this is under the assumption of good network condition and small input data size. In many cases, when the network condition is poor or for applications such as video analytics where a large amount of data is processed, offloading may take longer time, and thus may not be the best option. 

Recently, many companies such as Huawei, Qualcomm, and Samsung are developing dedicated {\em Neural Processing Units (NPUs)} for mobile devices, which can process AI features. With NPU, the running time of these deep learning models can be significantly reduced.
For example, the processing time of the ResNet-50 model \cite{he-iccv15} is about one second using CPU, but only takes about 50 $ms$ with NPU.
Although NPUs are limited to advanced phone models at this time, this technique has great potential to be applied to other mobile devices, and even for IoT devices in the future. 

There are some limitations with NPU. 
First, NPU uses 16 bits or 8 bits to represent the floating-point numbers instead of 32 bits in CPU. 
As a result, it runs CNN models much faster but less accurate compared to CPU. 
Second, NPU has its own memory space and sometimes the CNN models are too large to be loaded into memory.
Then, the CNN model has to be compressed in order to be loaded by NPU and then reducing the accuracy.
For instance, HUAWEI mate 10 pro only has 200MB memory space for NPU and many advanced CNN models must be compressed at the cost of accuracy.

There is a tradeoff between the offloading based approach and the NPU based approach. Offloading based approach has good accuracy, but may have longer delay under poor network condition. On the other hand, NPU based approach can be faster, but with less accuracy. 
In this paper, we propose FastVA, a framework that combines these two approaches for real time video analytics on mobile devices.
The major challenge is to determine when to offload the computation and when to use NPU based on the network condition, the video processing time, and the accuracy requirement of the application. 

Consider an example of a flying drone.
The camera on the drone is taking videos which are processed in real time to detect nearby objects to avoid crashing into a building or being trapped by a tree. 
To ensure no object is missed, the detection result should be as accurate as possible.
Here, the time constraint is critical and we should maximize the detection accuracy under such time constraint. 
For many other mobile applications such as unlocking a smartphone, making a payment through face recognition, or
using Google glasses to enhance user experience by recognizing the objects or landmarks and showing related information, accuracy and processing time are both important. Hence, we should achieve a better tradeoff between them. 

Based on the accuracy and the processing time requirement of the mobile application, we study two problems: 
\textit{Max-Accuracy} where the goal is to maximize the accuracy under some time constraints, and 
\textit{Max-Utility} where the goal is to maximize the utility which is a weighted function of accuracy and processing time. 
To solve these two problems, we have to determine when to offload the computation and when to use NPU. The solution depends on the network condition, the special characteristics of NPU, and the optimization goal. 
We will formulate them as integer programming problems, and propose heuristics based solutions. 

Our contributions are summarized as follows.
\begin {itemize}
\item
We study the benefits and limitations of using NPU to run CNN models to better understand the characteristics of NPU in mobile.

\item
We formulate the Max-Accuracy problem and propose a heuristic based solution. 

\item
We formulate the Max-Utility problem and propose an approximation based solution. 

\item
We implement FastVA on smartphones and compare it with other techniques through extensive evaluations.

\end {itemize}

The rest of the paper is organized as follows.
Section \ref{sec:related-work} presents related work.
Section \ref{sec:preliminary} studies the benefits and limitation of NPU and provides an overview of FastVA.
We formulate the \textit{Max-Accuracy Problem} and propose a solution in Section \ref{sec:max-acc}.
In Section \ref{sec:max-utl}, we study the \textit{Max-Utility Problem} and give a heuristic based solution.
Section \ref{sec:evaluation} presents the evaluation results and
Section \ref{sec:conclusion} concludes the paper.

\section{Related Work} \label{sec:related-work}

Over past years, there have been significant advances on object recognition with CNN models.
For example, GoogleNet \cite{szegedy-cvpr15} and ResNet \cite{he-cvpr16} can achieve high accuracy.
However, these CNN models are designed for machines with powerful CPU and GPU, and it is hard to run them on mobile devices due to limited memory space and computation power.
To address this issue, various model compression techniques have been developed.
For example, in \cite{bhattacharya-sensys16}, the authors propose to separate convolutional kernels from the convolutional layers and compress the fully-connected layers  to reduce the processing time of CNN models.
Liu \textit{et al.} \cite{liu-cvpr15} optimize the convolutional operations by reducing the redundant parameters in the neural network.
FastDeepIoT \cite{yao-sensys18} compresses the CNN models by optimizing the neural network configuration based on the non-linear relationship between the model architecture and the processing time.
Although the efficiency can be improved through these model compression techniques, the accuracy also drops. 

Offloading techniques have been widely used to address the resource limitation of mobile devices. 
MAUI \cite{cuervo-mobisys10} and many other works \cite{Geng-ICNP2015, Geng-infocom18, Geng-twc18} are general offloading frameworks that optimize the energy usage and the computation overhead for mobile applications.
However, these techniques have limitations when applied to video analytics where a large amount of video data has to be uploaded to the server. 
To reduce the data offloading time, some local processing techniques have been proposed to filter out the less important or redundant data.
For example, Glimpse \cite{chen-sensys15} only offloads a frame when the system detects that the scene changes significantly.
Similar to Glimpse, MARVEL \cite{chen-sensys18} utilizes the inertial sensors to detect and filter out the redundancy before offloading. However, both MARVEL and Glimpse may not work well when the network condition is poor or when there are many scene changes.
To address this issue, other researchers consider how to guarantee a strict time constraint by running different CNN models locally under different network conditions \cite{han-mobisys16, ran-infocom18}. 

Some recent work focuses on improving the execution efficiency of CNN models on mobile devices through hardware support. 
For example, Cappuccino \cite{motamedi-esl19} optimizes computation by exploiting imprecise computation on the mobile system-on-chip (SoC). Oskouei \textit{et al.} \cite{oskouei-mm16} developed an Android library called CNNdroid for running
CNN models on mobile GPU.
DeepMon \cite{huynh-mobisys17} leverages GPU for continuous vision analysis on mobile devices.
DeepX \cite{lane-ipsn16} divides the CNN models into different blocks which can be efficiently run on CPU or GPU.
The processing time is reduced by scheduling these blocks on different processors, such as CPU and GPU.
Different from them, we use NPUs.

\section{Preliminary} \label{sec:preliminary}

\begin{figure}[htb]
	\centering
	\subfloat[Processing Time Comparison]{\label{fig:NPU-process-time}\includegraphics[width=0.49\linewidth,valign=c]{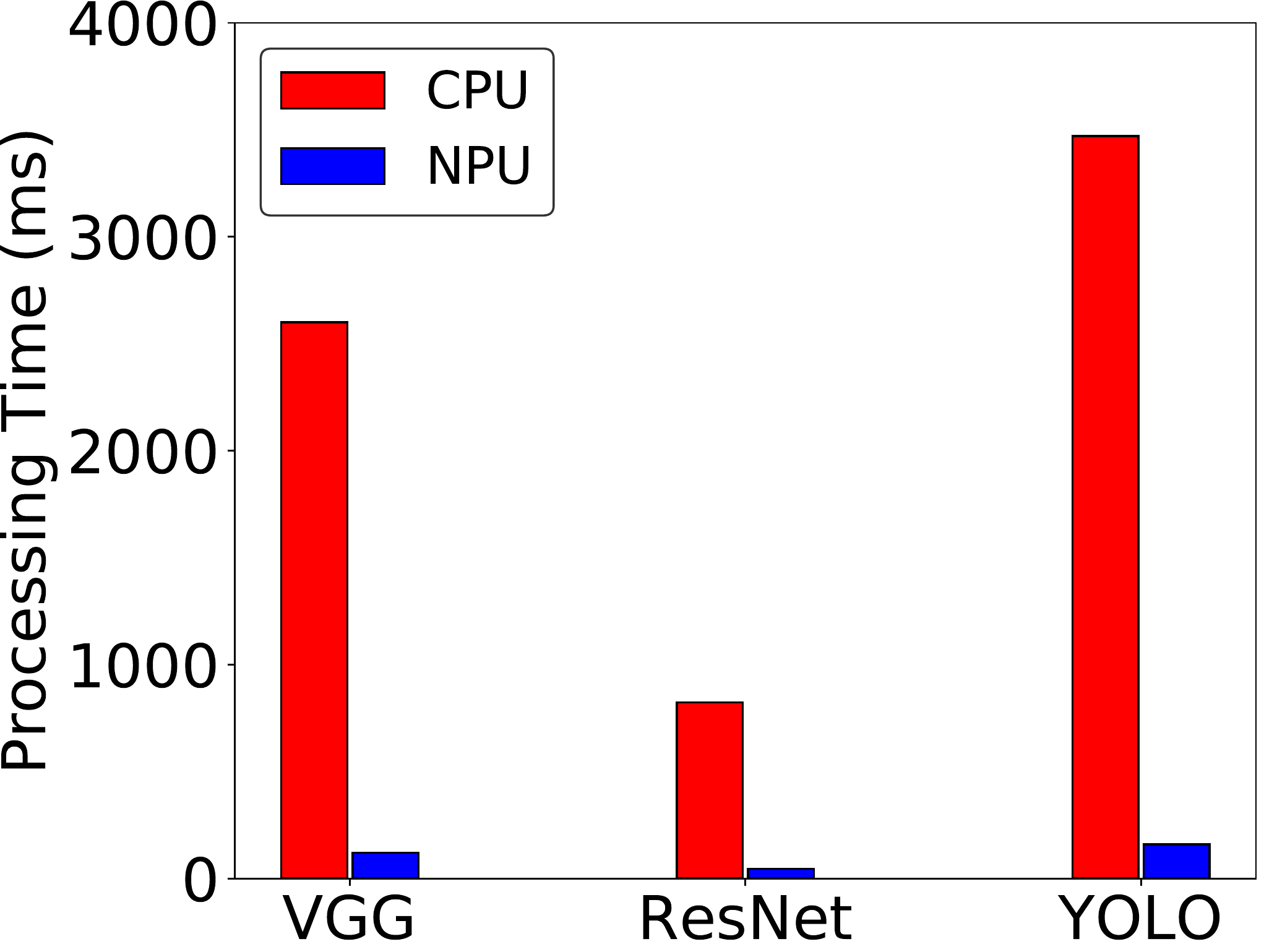}}
	\,
	\subfloat[Accuracy Comparison]{\label{fig:NPU-accuracy}\includegraphics[width=0.49\linewidth,valign=c]{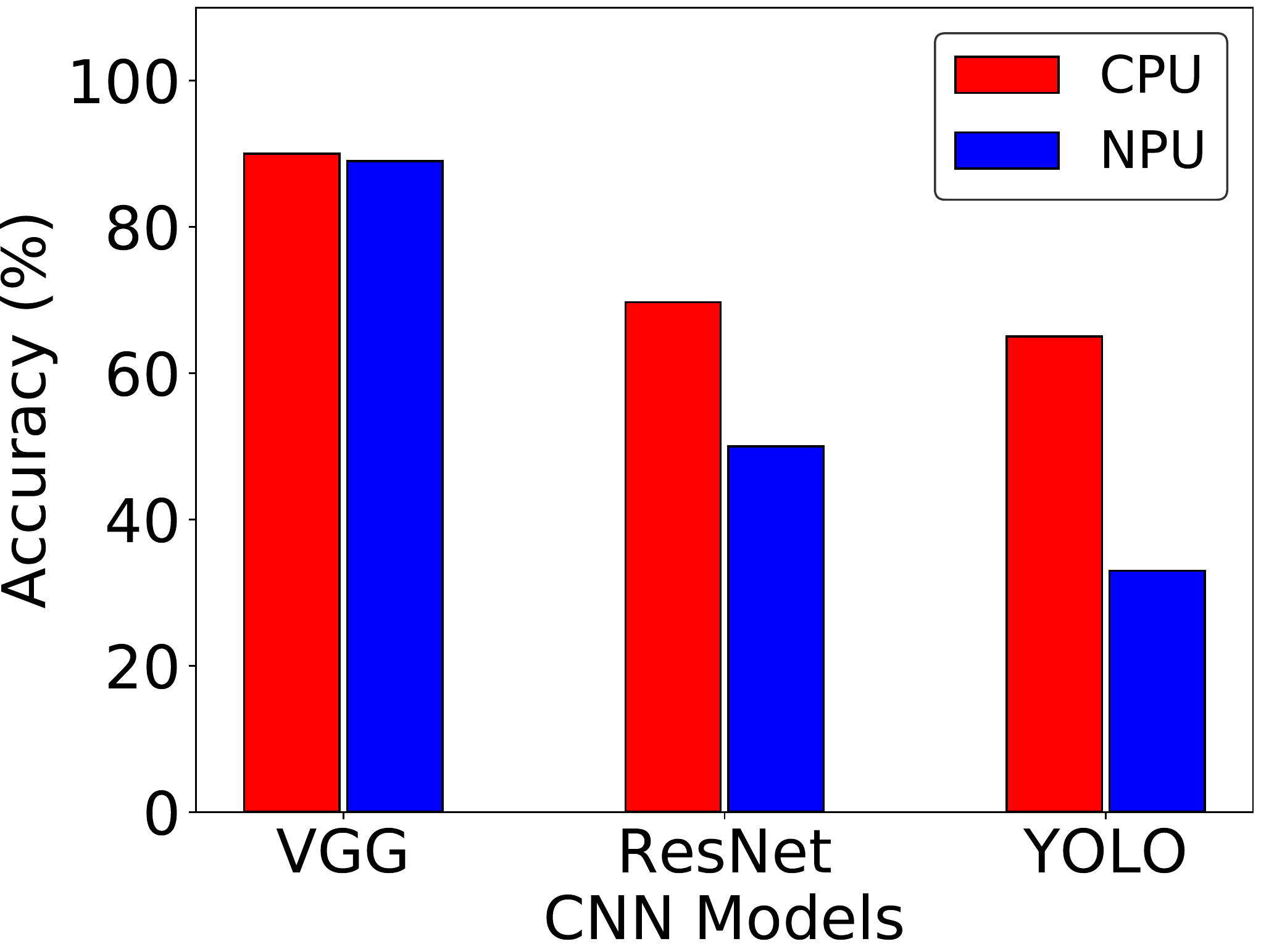}}
	\caption{Performance Comparison of NPU and CPU.}
	\label{fig:NPU-profile}
	\vspace{-1em}
\end{figure}

A video frame can be processed locally by the NPU or offloaded to the server. The decision depends on the application requirements on the tradeoff between accuracy and processing time. More importantly, the decision depends on how various CNN models perform on NPUs in terms of accuracy and processing time. In this section, we first show some results on how various CNN models perform on NPUs and local CPUs to understand the characteristics of NPU, and then give an overview of the proposed FastVA framework. 

\subsection{Understanding NPU}

\begin{figure*}
	\centering
	\subfloat[The Original Image]{\label{fig:detection-original-image}\includegraphics[width=0.24\linewidth,valign=c]{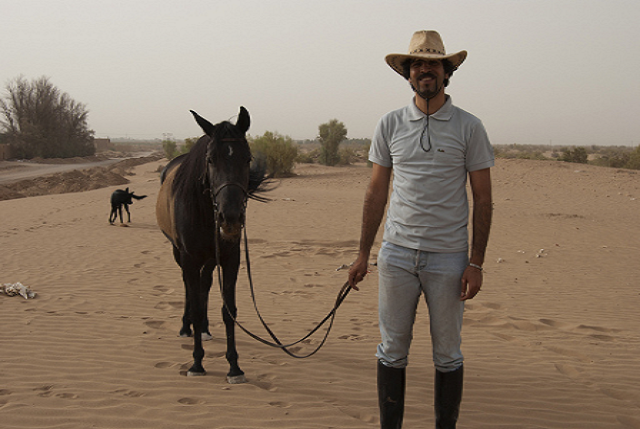}}
	\,
	\subfloat[Ground Truth]{\label{fig:detection-ground-truth}\includegraphics[width=0.24\linewidth,valign=c]{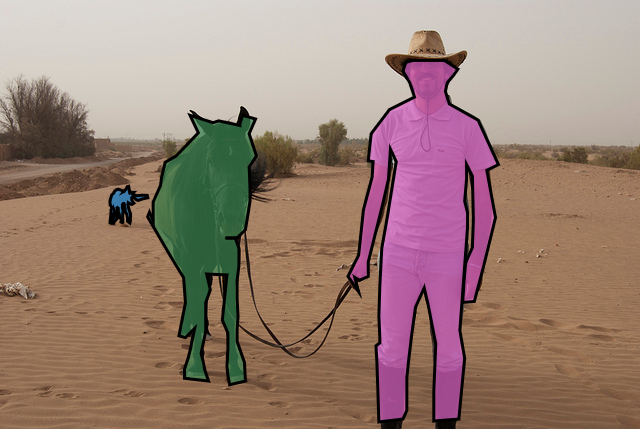}}
	\,
	\subfloat[Detection Result on CPU]{\label{fig:cpu-detect-result}\includegraphics[width=0.24\linewidth,valign=c]{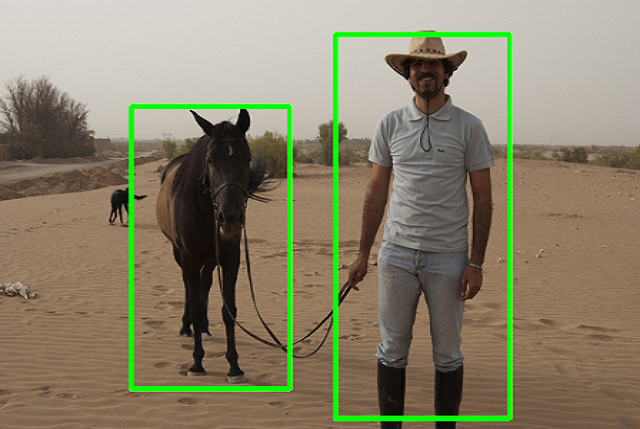}}
	\,
	\subfloat[Detection Result on NPU]{\label{fig:npu-detect-result}\includegraphics[width=0.24\linewidth,valign=c]{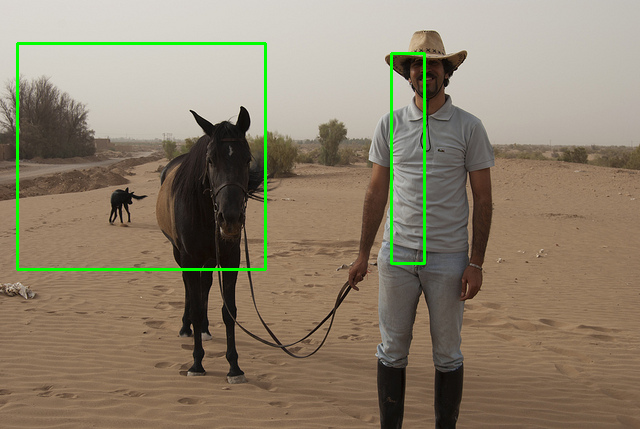}}
	\caption{Detection result with YOLO Small }
	\label{fig:detection-comparision}
	\vspace{-0.7cm}
\end{figure*}

\begin{figure}[htb]
	\centering
	\includegraphics[width=\linewidth,valign=c]{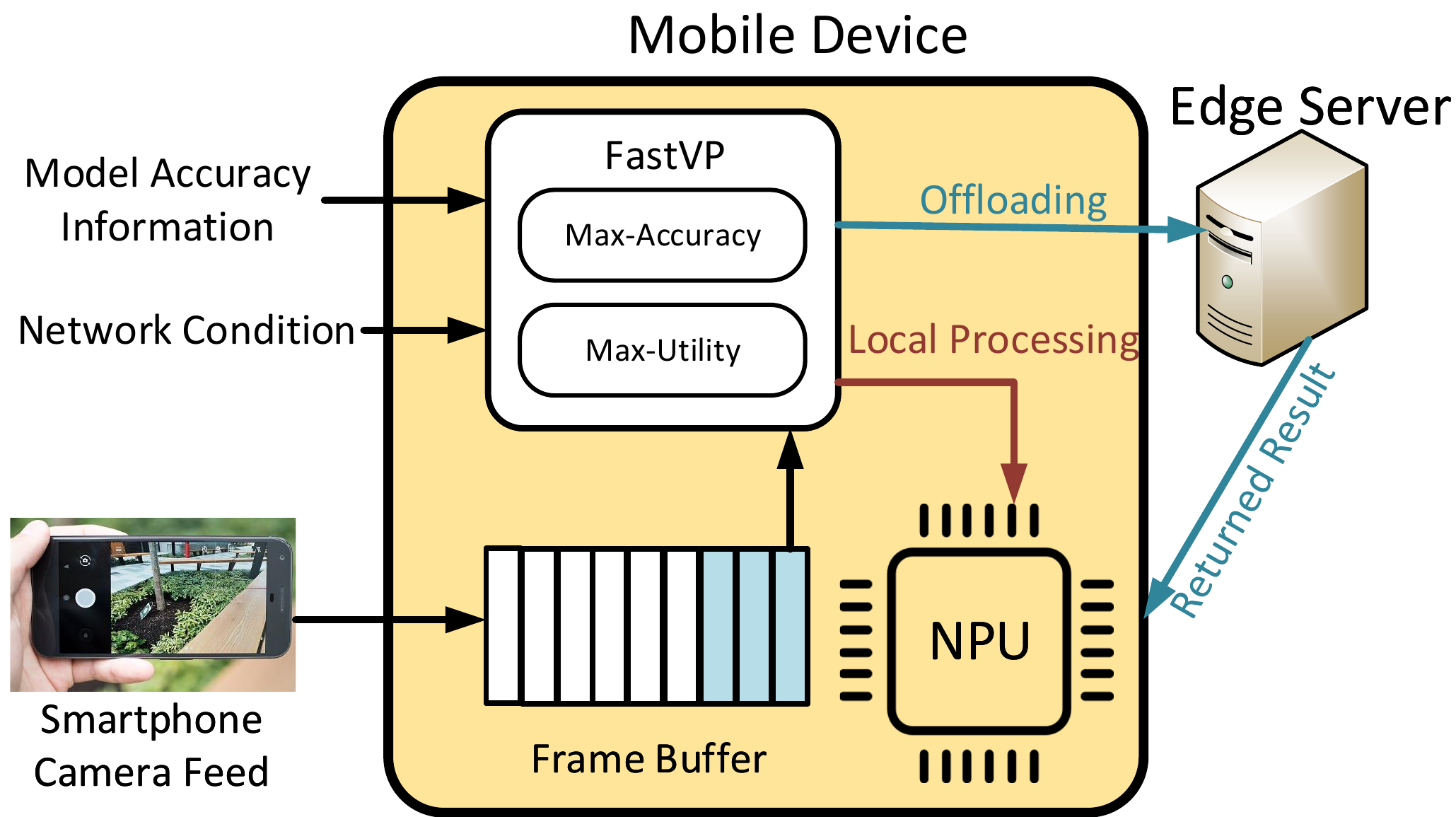}
	\caption{FastVA overview}
	\label{fig:overview}
	\vspace{-0.5cm}
\end{figure}

To have a better understanding of NPU, we compare the accuracy and the processing time of running different CNN models on NPU and CPU. 
The experiment is conducted on HUAWEI mate 10 pro which has a NPU, and the results are shown in Figure \ref{fig:NPU-profile}.
Three CNN models are used in the evaluations and the details are as follows. 

\begin{itemize}
	\item The VGG model \cite{parkhi-bmvc15} which is used for face recognition. 
	In the experiment, we use the face images from the LFW dataset \cite{lfw-dataset}. 
	
	\item The ResNet-50 model \cite{he-cvpr16} which is used for object recognition. 
	The evaluation was based on 4000 object images randomly chosen from the VOC dataset \cite{everingham-ijcv15}, and the results were based on the 
	Top-1 accuracy.
	
	\item The YOLO Small model \cite{redmon-cvpr16} which is designed for detecting objects in an image.
	The evaluation was based on 100 images randomly chosen from the COCO dataset \cite{lin-eccv14}, and the results were based on the F1-score.
	
\end{itemize}

As shown in Figure \ref{fig:NPU-profile}(a), compared to CPU, running VGG,  ResNet-50, or Yolo small on NPU can significantly reduce the processing time, by 95\%. As shown in Figure \ref{fig:NPU-profile}(b), the accuracy loss of using NPU is different for different CNN models.
For example, compared to CPU, using NPU has similar accuracy when running VGC, 20\% accuracy loss when running ResNet, and the F1-score drops to 0.3 when running YOLO Samll. This is because different CNN models have different ways to process these images. 

VGG compares the similarity between the two feature vectors $[x_1, x_2, \ldots, x_n]$ and $[y_1, y_2, \ldots, y_n]$, 
which are extracted from the face images.
They belong to the same person if the similarity is below a predefined threshold.
The similarity can be measured by the square of the Euclidean distance between two vectors, where $d = \sum_{i=1}^{n} (x_i - y_i)^2$.
Since NPU uses less number of bits to represent the floating point numbers, some errors will be introduced. 
The error introduced by NPU changes $d$ to $d' = \sum_{i=1}^{n} (x_i - y_i + \epsilon_i)^2$.
If we consider $\epsilon_i$ as noise data with mean value 0, the expected value of $d'$ is equal to $E(d) + \sum_{i} E(\epsilon_i^2)$.
Since $\epsilon_i$ is small, it will not change the relationship between $d$ and the threshold too much for most input data, and hence 
has the same level of accuracy as CPU.

Suppose the feature vector extracted from an image is $\Vec{f_1} = [x_1, x_2, \ldots x_n]$, ResNet-50 classifies the image based on the largest element.
Assume $x_p$ and $x_q$ are the two largest elements in $\Vec{f_1}$, and $x_p > x_q$. Then, the image will be classified as the $p^{th}$ category image.
The errors introduced by NPU may change the elements to $x'_p= x_p + \epsilon_1$ and $x'_q = x_q + \epsilon_2$.
If the difference between $x_p$ and $x_q$ is small, $x'_q$ may be larger than $x'_p$ due to errors $\epsilon_1, \epsilon_2$.
Then, the object will be classified as the $q^{th}$ category, and getting a wrong result. 

YOLO Small is much more complex than ResNet and VGG.
Its feature vector includes information related to location, category and size of the objects, and a small error in the feature vector can change the result.
For instance, Figure \ref{fig:detection-comparision} shows the detection result of a test image using CPU and NPU.
Figure \ref{fig:detection-comparision}(c) shows the result with CPU, where the bounding boxes accurately include the objects, and the objects can be correctly recognized as horse and person respectively.
Figure \ref{fig:detection-comparision}(d) shows the result with NPU. As can be seen, the detection result only includes part of the objects. The center and the size of the objects are incorrect, leading to wrong detection results. 

From these evaluations, we can see that NPU runs much faster than CPU; however, it may not always be the best choice for running CNN models, especially when accuracy is important.

\subsection{FastVA Overview}

The overview of FastVA is shown in Figure \ref{fig:overview}.
FastVA can process frames through offloading or local processing with NPU.
For offloading, FastVA may reduce the frame resolution before transmitting so that more frames can be uploaded at the cost of accuracy.
Similarly, multiple CNN models can be used, where a smaller CNN model can reduce the processing time at the cost of accuracy, and 
a larger model can increase the accuracy at the cost of longer processing time.
If several CNN models are available, FastVA will choose the proper one for processing under different constraints.
Based on the network condition and the accuracy of the CNN models, 
the schedule decision will be determined by the proposed Max-Accuracy or Max-Utility. 
To provide real time video analytics, FastVA ensures that the processing of each video frame is completed within a time constraint.

Due to the limited computational and bandwidth resources, the scheduling needs to be designed carefully in order to maximize the accuracy or utility.
In the following sections, we formulate and solve the Max-Accuracy problem and the Max-Utility problem.

\section{The Max-Accuracy Problem} \label{sec:max-acc}

In this section, we study the Max-Accuracy problem which aims to maximize the accuracy under some time constraints. 
We first formulate the problem and then propose a heuristic based solution. 

\subsection{Problem Formulation}

\begin{table}[ht]
	\centering
	\begin{tabular}{|c | l |} 
		\hline
		Notation & Description \\ \hline
		$I_i$ & the $i$th frame  \\ \hline
		$S(I_i, r)$ & the data size of the frame $I_i$ in resolution $r$ \\ \hline
		$T^{\textit{npu}}_j$ & The processing time of $j^{th}$ model on NPU \\ \hline
		$T^{\textit{o}}_j$ & Processing time using the $j^{th}$ model on the server \\ \hline
		$a(j, r)$ & The accuracy of the $j^{th}$ model with input images  \\
		&  in resolution $r$\\ \hline
		$T_c$ & Communication delay between mobile device and server \\ \hline
		$B$   & upload bandwidth (data rate) \\ \hline
		$f$   & video frame rate (fps) \\ \hline
		$\gamma$ & the time interval between two consecutive frames \\ \hline
		$T$   & the time constraint for each frame\\ \hline
		$n$   & the number of video frames that needs to be processed \\ \hline
	\end{tabular}
	\caption{Notation.}
	\label{table:notations}
	\vspace{-0.5cm}
\end{table}

Assume the incoming frame rate is $f$, the time interval between two consecutive video frames is $\gamma = \frac{1}{f}$.
For the $i^{th}$ frame in the video, assume its arrival time is $i \gamma$ and FastVA needs to process it before $T + i \gamma$, where $T$ is the time constraint.
For each frame, it can either be processed locally by the NPU or offloaded to the server.
Multiple CNN models are used on the edge server and the mobile device to process these frames.
If the frame $I_i$ is processed by the $j^{th}$ model locally, the corresponding processing time is $T^{\textit{npu}}_j$.
If $I_i$ is processed at the edge server, the data can be offloaded with the original resolution or reduce the resolution to $r$ before uploading to save bandwidth.
Let $B$ denote the upload bandwidth and let $T_c$ denote the communication delay between the edge server and the mobile device.
Then, it takes $\frac{S(I_i, r)}{B} + T^{\textit{o}}_j + T_c$ to transmit the $i^{th}$ frame in resolution $r$ and receive the result from the server.
Although the transmission time can be reduced by reducing the frame to a lower resolution, the accuracy is lower.

The notations used in the problem formulation and the algorithm design are shown in Table \ref{table:notations}.
The Max-Accuracy problem can be formulated as an integer programming in the following way.

\begingroup
\allowdisplaybreaks
\begin{align}
\max \quad & \frac{1}{n} \displaystyle\sum^{n}_{i=0} \sum_j \sum_r a(j, r) X^j_i Y^r_i \\
\textrm{s.t.} \quad 
&  \sum^{i}_{k=i'} \sum_j T^{\textit{npu}}_j X^j_k  + i' \gamma  \leq T + i \gamma, \, \forall i, i', i' \leq i \\
& D(i', i) + T_c \leq i \gamma + T,  \, \forall i, i', i' \leq i \\
& \sum_j X^j_i = 1, \,\forall i \\
& \sum_r Y^r_i = 1, \,\forall i \\
& Y^r_i, X^j_i \in \{0, 1\} \,\forall i, j
\end{align}
\endgroup
Where $D(i', i) = i' \gamma + \sum_j(\sum_r \sum^{i}_{k=i'} \frac{S(k, r) Y^r_k}{B} + T^o_j X^j_i)$ is the offloading time for the frames that arrive between $I_{i'}$ and $I_i$. $X^j_i$ is a variable to show which model is used to process the frame and $Y^r_i$ is a variable to show which resolution the frame is resized to before offloading.
If $X^j_i = 0$, the frame $I_i$ is not processed by the $j^{th}$ model.
If $X^j_i = 1$, the frame $I_i$ is run by the $j^{th}$ model.
If $Y^r_i = 1$, the frame $I_i$ is resized to resolution $r$ before offloading.

Objective (1) is to maximize the accuracy of the processed frames in the time window.
Constraint (2) specifies that all local processed frames should be completed before the deadline, 
and constraint (3) specifies that the results of the offloaded frames should be returned within the time constraint. 

\subsection{Max-Accuracy Algorithm}

\begin{algorithm}[ht]
	\SetAlgoLined
	\KwData{Video frames in the buffer}
	\KwResult{Scheduling decision}
	\SetKw{Continue}{continue}
	\SetKw{Break}{break}
	\SetKw{Return}{return}
	\DontPrintSemicolon
	
	The frame schedule list $S \leftarrow \{\}$ \\
	$A \leftarrow 0$, $n_s \leftarrow$ the number of models on the server side \\
	\For{each possible resolution $r$} {
		Resize the $I_0$ to the resolution $r$ \\
		$A' \leftarrow 0, S' \leftarrow \{\}$ \\
		Sort the remote models in the descending order based on their accuracy $a(j, r)$. \\
		\For{$j$ from 1 to $n_s$} {
			\If{$t + T^{\textit{o}}_j + T_c \leq T$} {
				Add $(0, j, r)$ to $S'$ \\
				$A' \leftarrow A' + a(j, r)$ \\
				\Break \\
			}
		}
		
		$n_l \leftarrow  \floor{\frac{S(I_i, r)}{B \gamma}}$ \\
		
		Compute $H(i, t)$ according to Equation \ref{eq:local-initcase} and \ref{eq:local-phase} for $i \in [1, n_l]$ and $t \in [\gamma, n_l \gamma + T]$ \\
		$h' \leftarrow \max_{t} H(n_l, t)$  \\
		$t' \leftarrow \arg \max_{t} H(n_l, t)$ \\
		
		\For{$i$ from $n_l$ to $1$} {
			\For{each local model $j$}{
				\If {$H(i-1, t' - T^{\textit{npu}}_j) = h'$} {
					Add $(i, j, r_{max})$ to $S'$ \\
					$t' \leftarrow t' - T^{\textit{npu}}_j, h' \leftarrow h' - a(j, r_{max}), A' \leftarrow A' + a(i, j)$ \\
					\Break \\
				}
			}
		}
		\If{$\frac{A'}{n_l + 1} > A$} {
			$A \leftarrow \frac{A'}{n_l + 1}, S \leftarrow S'$ \\
		}
	}
	\Return $S$ \\
	
	\caption{Max-Accuracy Algorithm}
	\label{alg:max-acc}
\end{algorithm}

A brute force method to solve the Max Accuracy Problem is to try all the possible scheduling options, and 
it takes $O((n_c * n_r)^{n})$, where $n_c$ is the number of CNN models available for processing the frames and $n_r$ is the number of resolution options. 
Since the brute force method is impractical, we propose a heuristic solution.
The basic idea is as follows. 
Since offloading based approach can achieve better accuracy than NPU based approach for the same CNN model,
the arriving video frame should be offloaded as long as there is available bandwidth. 
Due to limited bandwidth, some frames cannot be offloaded and will be processed by the NPU locally.
More specifically, our Max-Accuracy algorithm consists of multiple rounds.
In each round, there are two phases: offload scheduling phase and local scheduling phase.  
In both phases, the right CNN model is selected to process the video frame within the time constraint and maximize the accuracy.

\subsubsection{Offload Scheduling}

In this phase, the goal is to find out the CNN model that can be used for processing the offloaded video frame within time constraint and maximize the accuracy. 
Assume that the network interface is idle and $I_0$ is the new frame arriving at the buffer.
$I_0$ will be resized to resolution $r$ and offloaded to the server.
On the server side, the only requirement for selecting the CNN model is the time constraint, which requires the result must be returned in time.
In other words, the constraint $\frac{S(I_0, r)}{B} + T_c + T^{\textit{o}}_j \leq T$ must be satisfied for the uploaded frame $I_0$.
A CNN model will be selected if it can satisfy the time constraint and has the highest accuracy on images with resolution $r$.
Since video frames arrive at a certain interval $\gamma$, $n_l = \floor{\frac{S(I_0, r)}{B \gamma}}$ frames will be buffered 
while $I_0$ is being transmitted. 
These frames will be processed locally, and the local scheduling phase will be used to determine their optimal scheduling decision. 

\subsubsection{Local Scheduling}

In this phase, the goal is to find out the CNN model that can be used for processing the video frame within time constraint and maximize the accuracy. 
For each CNN model, the video processing time and the accuracy vary. A simple dynamic programming algorithm is used to find an optimal scheduling decision.
More specifically, let $H(k, t)$ denote the optimal accuracy for processing the first $k$ frames with time constraint $t$, where $k \in [0, n_l]$.
Then, frame $I_1$ arrives at the frame buffer at time $\gamma$ and the last frame $I_{n_l}$ must be processed before time $n_l \gamma + T$, thus $t \in [\gamma, n_l \gamma + T]$.
If it is impossible to process all $k$ frames within $t$, $H(k, t) = - \infty$. Initially, since the frame $I_0$ is offloaded to the server, $H(0, t)$ can be computed as follow:

\begin{equation} \label{eq:local-initcase}
H(0, t) = \begin{cases}
-\infty, & \text{if $t < T^{idle}$} \\
0 &  \textrm{Otherwise}
\end{cases}
\end{equation}

{\noindent}where $T^{idle}$ is the queuing time for $I_1$.

For frame $I_k (k > 0)$, it can be processed on one of the local CNN model $j$. $H(k, t)$ can be computed as follow:

\begin{equation} \label{eq:local-phase}
H(k, t) = \begin{cases}
-\infty, & \hspace{-1.75cm} \text{if $\forall j, k \gamma + T^{\textit{npu}}_j < t$} \\
\max_j (H(k-1, t-T^{\textit{npu}}_j) + a(j, r_{max})), &  \hspace{-0.3cm} \textrm{Otherwise}
\end{cases}
\end{equation}

Based on the computed $H(k, t)$, the scheduling decision can be made by backtracking.
The Max-Accuracy algorithm is summarized in Algorithm \ref{alg:max-acc}.
Lines 4-11 are the offloading scheduling phase and Lines 12-21 are for the local scheduling phase.
In the algorithm, a variable $A$ is used for tracking the maximum accuracy that is found so far, and its corresponding schedule decision is maintained in $S'$.
The frame schedule list $S'$ is a list of pair $(i, j, r)$, which means that frame $I_i$ is processed by the $j^{th}$ model with resolution $r$.
The running time of our algorithm is $O(n_r * n_c * n)$

\section{Max-Utility Problem} \label{sec:max-utl}

In this section, we study the Max-Utility problem. 
The goal is to maximize the utility which is a weighted function of accuracy and video processing time.
We first formulate the problem and then propose an approximated based solution.

\subsection{Problem Formulation}

With time constraint, FastVA may not be able to process all frames using the CNN model with the highest accuracy. To achieve high accuracy with limited resources, FastVA may skip some frames whose queuing time is already close to its time constraint. 
With the notations used in the last section, the length of the video is $n \gamma$ and $\sum_{i} \sum_{j} X^{j}_{i}$ is the total number of frames to be processed. Then, the video is processed at a real frame rate of $\frac{\sum_{i} \sum_{j} X^{j}_{i}}{n \gamma}$.
The average accuracy can be computed as $\frac{\sum_{i} \sum_{j} a(j, r)X^{j}_{i} Y^{r}_{i}}{\sum_{i} \sum_{j} X^{j}_{i}}$.
Let $\alpha$ denote the tradeoff parameter between accuracy and processing time (measured with the frame processing rate).
Then, the utility can be computed as $\sum_{i} \sum_{j} \frac{X^{j}_{i}}{n \gamma} + \alpha \frac{\sum_{i} \sum_{j} a(j, r)X^{j}_{i} Y^{r}_{i}}{\sum_{i} \sum_{j} X^{j}_{i}}$.

Similar to the Max-Accuracy problem, the Max-Utility Problem can be formulated as an integer programming in the following way.

\begingroup
\allowdisplaybreaks
\begin{align}
\max \quad &  \sum^{n}_{i=0} \sum_{j} \frac{X^{j}_{i}}{n \gamma} + \alpha \frac{\sum_{i} \sum_{j} a(j, r)X^{j}_{i} Y^{r}_{i}}{\sum_{i} \sum_{j} X^{j}_{i}} \\ 
\textrm{s.t.} \quad 
& \sum^{i}_{k=i'} \sum_j T^{\textit{npu}}_j X^j_k + i' \gamma  \leq i \gamma + T, \, \forall i, i', i' \leq i \\
& D(i', i) + T_c \leq i \gamma + T,  \, \forall i, i', i' \leq i \\
& \sum_j X^j_i \leq 1, \,\forall i \\
& \sum_r Y^r_i \leq \sum_j X^j_i, \,\forall i \\
& Y^r_i, X^j_i \in \{0, 1\}, \,\forall i, j
\end{align}
\endgroup

Objective (9) maximizes the utility. 
Constraints (10) and (11) specify that the frames must be processed within the time requirement.
Constraint (13) specifies that each frame can at most be processed by a CNN model either remotely or locally.
Constraint (14) specifies that each image can only be resized to a certain resolution.

\subsection{Max-Utility Algorithm}

Since the problem is NP-hard, it costs too much time to find the optimal solution.
Therefore, we propose a heuristic based algorithm (called the Max-Utility Algorithm) to solve it. 
The basic idea of the algorithm is as follows.
Since the offloading based approach can achieve better accuracy than NPU based approach for the same CNN model, our algorithm first maximizes the utility by offloading the arriving video frame with the available bandwidth.
Due to the limited bandwidth, some frames will not be offloaded and our Max-Utility algorithm further improves the utility using a dynamic programming algorithm to decide which frames should be skipped and which frames should be processed locally.

Assume the network interface is idle when a new frame $I_0$ arrives in the buffer.
$I_0$ will be resized to a resolution $r$ and offloaded to the server.
The offloading time for this frame is $\frac{S(I_0, r)}{B}$, which means the frames are offloaded at the frame rate $\frac{B}{S(I_0, r)}$.
The schedule decision for $I_0$ is made by solving $\max_{r, j} \frac{B}{S(I_0, r)} + \alpha \times a(j, r)$.
Since the result from the server should be received within the time limitation,
the constraint $T \geq \frac{S(I_0, r)}{B} + T_c + T^{\textit{o}}_j$ should be satisfied.
$n_l = \floor{\frac{S(I_0, r)}{B}}$ frames will be buffered while $I_{0}$ is being transmitted.
These frames will be processed locally, and a dynamic programming algorithm is used to find out the optimal solution.

In the algorithm, an array $U(k)$ ($k \in [0, n_l]$) is maintained to find the schedule for maximizing the utility.  
$U(k)$ is a list of triples, and each triple is denoted as $(t, u, m)$, where utility $u$ is gained by processing $m$ out of the $k$ frames locally 
within time $t$.
Notice that not all possible triples are maintained in $U(k)$, and only the most efficient ones (i.e., with more utility and less processing time) are kept
More specifically, a triple $(t', u', m')$ is said to dominate another triple $(t, u, m)$ if and only if $t' \leq t, u'\geq u$.
Obviously, triple $(t', u', m')$ is more efficient than triple $(t, u, m)$ and all dominated triples will be removed from the list of $U(k)$.
Assume that $T^{idle}$ is the queuing time for $I_1$. Initially, $U(0) = \{(T^{idle}, 0, 0)\}$.
To add triples to the list of $U(k)$, we consider two cases: no processing, local processing.

\textbf{No processing:} 
In this case, the $k^{th}$ frame will not be processed.
Processing more frames may require a faster local CNN model to be used for processing.
In such cases, the average accuracy decreases and the utility may also decrease.
A better solution is to skip this frame.
Therefore, we will add all the triples in $U(k-1)$ to $U(k)$.

\textbf{Local Processing:} 
In this case, it requires $T^{\textit{npu}}_j$ time to process the frame using the $j^{th}$ model locally.
Since the frame does not need to be resized, it is processed with the maximum resolution $r_{max}$.
The new average accuracy can be computed as $A = \frac{m}{m+1} (u - \frac{m}{n_l \gamma}) + \alpha \frac{a(j, r_{max})}{m+1}$.
For each triple $(t, u, m) \in U(k-1)$, a new triple $(\max(t, k \gamma) + T^{\textit{npu}}_j, A + \frac{m + 1}{n_l \gamma}, m+1)$ is added to the list of $U(k)$.
Notice that all local processed frames should be finished within the time constraint. 
Therefore, $\max(t, k \gamma) + T^{\textit{npu}}_j \leq k \gamma + T$ should be satisfied for all new triples.

With the list of $U(k)$, we can find a schedule to maximize the utility.
The complete description of our algorithm is shown in Algorithm \ref{alg:dp-max-utility}.
In Lines 2-8, the algorithm maximizes the utility for the offloaded frame, and the schedule decision is determined for the local processing frames in Lines 9-27.
The running time of the algorithm is $O(n^2 * n_c)$.

\begin{algorithm}[ht]
	\SetAlgoLined
	\KwData{Video frames in the buffer}
	\KwResult{Scheduling decisions}
	\SetKw{Break}{break}
	\DontPrintSemicolon
	
	The frame schedule list $S \leftarrow \{\}$ \\
	$u \leftarrow 0$ \\
	\For{$j$ from 1 to $n_s$} {
		\For{each possible resolution $r$} {
			$u' \leftarrow \frac{B}{S(I_0, r)} + \alpha \times a(j, r)$ \\
			\If{$\frac{S(I_0, r)}{B} + T^{\textit{o}}_j + T_c \leq T$ and $u < u'$} {
				$p \leftarrow (0, j, r), u \leftarrow u'$
			}
		}
	}
	Add $p$ to $S$ \\

	$n_l \leftarrow \floor{\frac{S(I_0, r)}{B}}$, $U(0) \leftarrow \{(T^{idle}, 0, 0)\}$ \\
	
	\For{$i \leftarrow 1$ to $n_l$}{
		\For{each $(t, u, m) \in U(i-1)$}{
			Add $(t, u, m)$ to $U(i)$ \\
			\For{each local model j}{
				$t' \leftarrow \max(t, i \gamma) + T^{\textit{npu}}_j$ \\
				\If{$t' \leq T + i \gamma$ and $i \gamma < t$}{
					$A \leftarrow \frac{m}{m+1} (u - \frac{m}{n_l \gamma}) + \alpha \frac{a(j, r_{max})}{m+1}$ \\
					Add $(t', A + \frac{m + 1}{n_l \gamma}, m+1)$ to $U(i)$ \\
				}
			}
		}
		Remove the dominated pairs from U(i) \\
	}
	
	$(t', u', m') \leftarrow \arg \max_{(t, u, m) \in U(n_l)} u$ \\
	\For{$i$ from $n_l - 1$ to $0$}{
		\For{each pair $(t, u, m)$ in $U(i)$} {
			\For{each local model j}{
				$A \leftarrow \frac{m}{m'} (u - \frac{m}{n_l \gamma}) + \alpha \frac{a(j, r_{max})}{m'}$ \\
				\If{$t + T^{\textit{npu}}_j = t'$ and $A + \frac{m'}{n_l \gamma} = u'$}{
					Add $(i+1, j, r_{max})$ to $S$ \\
					$t' \leftarrow t, u' \leftarrow u, m' \leftarrow m$ \\
					\Break \\
				}
			}
		}
	}

	\Return $S$ \\
	\caption{Max-Utility Algorithm}
	\label{alg:dp-max-utility}
\end{algorithm}

\section{Performance Evaluations} \label{sec:evaluation}

In this section, we evaluate the performance of the proposed algorithms, Max-Accuracy and Max-Utility, and compare them with other approaches.

\subsection{Experiment Setup}

Currently, there are only a few smartphones on the market with dedicated NPUs.
In the evaluation, we use HUAWEI Mate 10 pro smartphone because it is equipped with NPU and it has a published HUAWEI DDK \cite{hiai} for developers.
Since NPU has a different architecture from CPU, the existing CNN models have to be optimized before running on NPU.
The HUAWEI DDK includes toolsets to do such optimization for NPU from CNN models trained by the deep learning frameworks Caffe \cite{jia-mm14}.
The HUAWEI DDK also includes the APIs to run the CNN models, and a few Java Native Interface (JNI) functions are provided to use the APIs on Android.
Since these JNI functions are hard coded for running a specific model, we have implemented more flexible JNI functions which can run different CNN models.

In FastVA, the frames are offloaded in the lossless PNG format.
The edge server is a desktop with AMD Ryzen 7 1700 CPU, GeForce GTX1070 Ti graphics card and 16 GB RAM.
We have installed the Caffe framework to run the CNN models on GPU.

In the experiment, object classifications are performed on mobile devices, which are common computer vision tasks for many mobile applications.
In our experiment, two different object recognition CNN models are used, ResNet-50 \cite{he-cvpr16} and SqueezeNet \cite{squeezenet}, which are 
well known and are widely used. Moreover, SqueezeNet has a compact structure and it is much smaller than ResNet.
It can be considered as a compressed model that runs faster than ResNet at the cost of accuracy. This allows the application to achieve tradeoffs between accuracy and processing time under different network condition and time constraint. 

In the evaluation, we use a subset of videos from the FCVID dataset \cite{jiang-TPAMI18}, which includes many real-world videos.
These videos have been used for training models related to object classification and activity recognition.
In our experiment, we focus on object classification, and thus activity recognition clips are not used.
Since the dataset is very large, about 1.9 TB, we randomly select 40 videos from the dataset and filter out the noisy data. 
Since the labels of FCVID and ImageNet are different, we map the labels produced by the CNN models to that used by the FCVID dataset.

We evaluate the proposed algorithms with different frame rates.
Most videos in the dataset use 30 fps, and thus we have to change their frame rate by decoding/encoding.
For both CNN models (ResNet-50 and SqueezeNet), the maximum resolution of the input image is 224x224 pixels.  
This resolution can be downsized for some offloading images, and we consider 5 different resolutions: 45x45, 90x90, 134x134,  179x179 and 224x224 pixels.
The time constraint for each frame is set to be 200 ms in all the experiments.
The running time of Max-Accuracy and Max-Utility algorithm is less than 1 ms on the smartphone and it is negligible compared to the time constraint (100 ms level).

\subsection{The Effects of CNN models}

\begin{table*}
	\begin{center}
		\begin{tabular}{|c|c|c|c|c|l|}
			\hline
			\multicolumn{2}{|c|}{CNN model}      & Processing Time (ms) & Transmission Time (ms) & \multicolumn{2}{c|}{Top-1 Accuracy} \\ \hline
			\multirow{2}{*}{ResNet}     & Local  & 52                   & 0                      & \multicolumn{2}{c|}{0.52}           \\ \cline{2-6} 
			& Server & 69                   & 39 - 242                    & \multicolumn{2}{c|}{0.67}           \\ \hline
			\multirow{2}{*}{SqueezeNet} & Local  & 17                   & 0                      & \multicolumn{2}{c|}{0.41}           \\ \cline{2-6} 
			& Server & 9                    & 39 - 242                    & \multicolumn{2}{c|}{0.51}            \\ \hline
		\end{tabular}
	\end{center}
	\vspace{-0.3cm}
	\caption{The performance of the CNN models.}
	\label{table:model-performance}
	\vspace{-0.8cm}
\end{table*}

\begin{figure}[t]
	\centering
	\includegraphics[width=0.49\linewidth,valign=t]{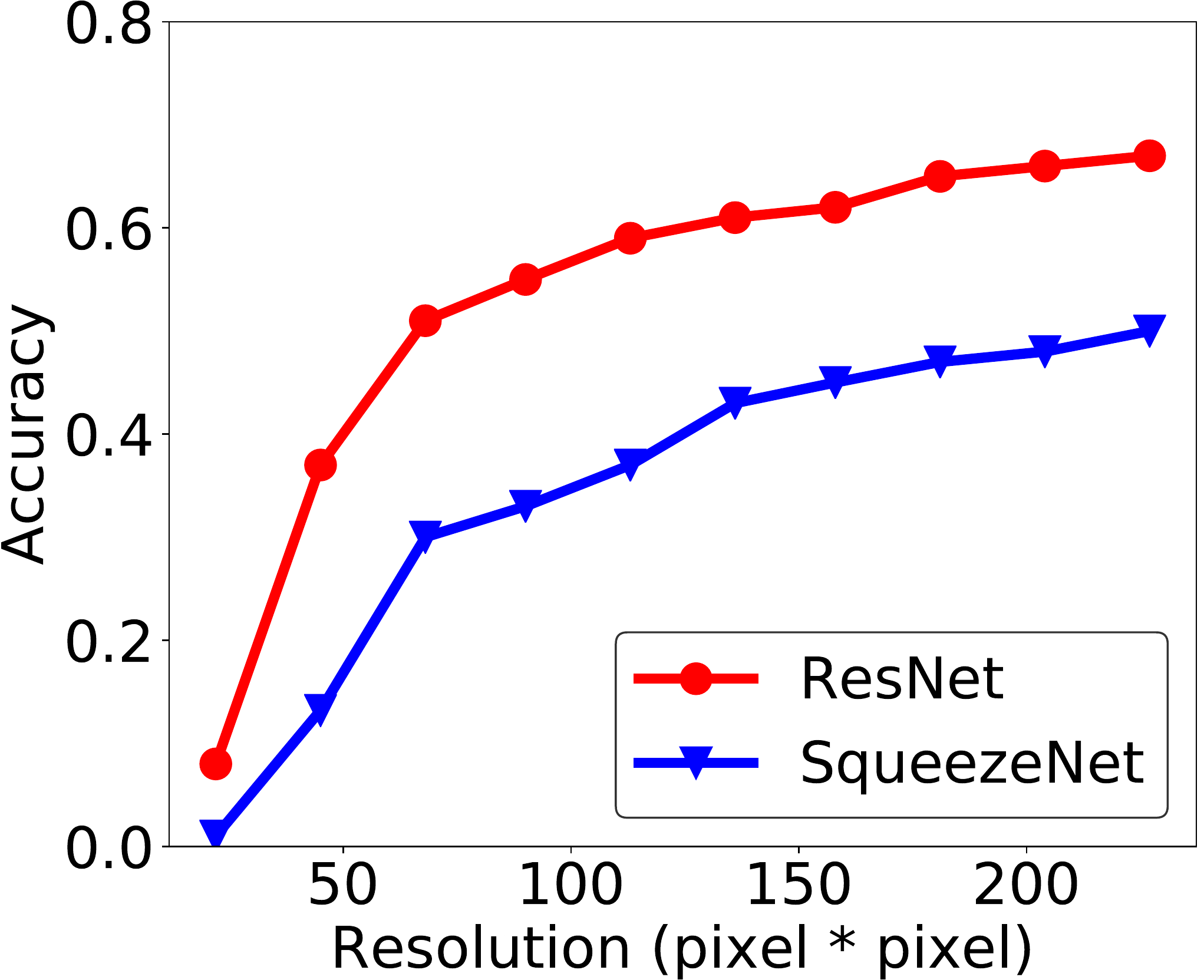}
	\caption{Accuracy vs. Resolution.}
	\label{fig:resolutions}
	\vspace{-0.5cm}
\end{figure}

To have a better understanding of how the CNN models perform, 
we run ResNet-50 and SqueezeNet on the edge server and NPU with randomly selected 4000 images from the VOC dataset. 
As shown in table \ref{table:model-performance}, the accuracy of ResNet-50 is about 30\% better than SqueezeNet on the server 
and it is about 25\% better than SqueezeNet on the NPU.
However, SqueezeNet is 700\% faster than ResNet-50 on the server and 
it is 300\% faster than ResNet-50 on the NPU.
Although running these CNN models has high accuracy on the server, there is a communication delay between the server and mobile device. 
As shown in the table, the transmission time can range from tens of milliseconds to hundreds of milliseconds 
based on the network condition and frame data size. When the network condition is poor, offloading may take much longer time than running on NPU.

Figure \ref{fig:resolutions} shows the tradeoff between accuracy and resolution.
We note that the accuracy does not scale linearly with the resolution.
The data in Table \ref{table:model-performance} and Figure \ref{fig:resolutions} are used for making scheduling decisions in FastVA.

\subsection{The Performance of Max-Accuracy}
	We compare the performance of Max-Accuracy with the following schedule algorithms.
	
	\begin{itemize}
		\item \textbf{Offload:} In this method, all frames must be offloaded to the edge server for processing.
		Each frame will be resized to a resolution so that it can be offloaded before the next frame arrives, and the server chooses the most accurate model that can process the frames and return the result within the time constraint.
		
		\item \textbf{Local:} 
		In this method, all frames are processed locally.
		It uses the proposed dynamic programming technique to find the optimal schedule decision for local processing.
		
		\item \textbf{DeepDecision:}
		This is a simplified version of DeepDecision \cite{ran-infocom18} which optimizes the accuracy and utility within the time constraint.
	    DeepDecision divides time into windows of equal size.
	    At the beginning of each time window, it picks a specific resolution and CNN model to process all the frames within a time window. 
		
		\item \textbf{Optimal:} This shows the upper bound for all methods.
		It tries all possible combinations and chooses the schedule that maximizes the accuracy.
		Notice that this method cannot be used for processing videos in real time since it takes too much time to search all possible schedules. 
		We can only find the optimal solution offline by replaying the data trace.
	\end{itemize}
	
	The performance of the schedule algorithms depends on several factors, the bandwidth, delay and the processing time requirement specified by the applications.

	\begin{figure}
		\centering
		\vspace{-0.3cm}
		\subfloat[Frame Rate 30 fps]{{
				\includegraphics[width=0.49\linewidth, valign=t]{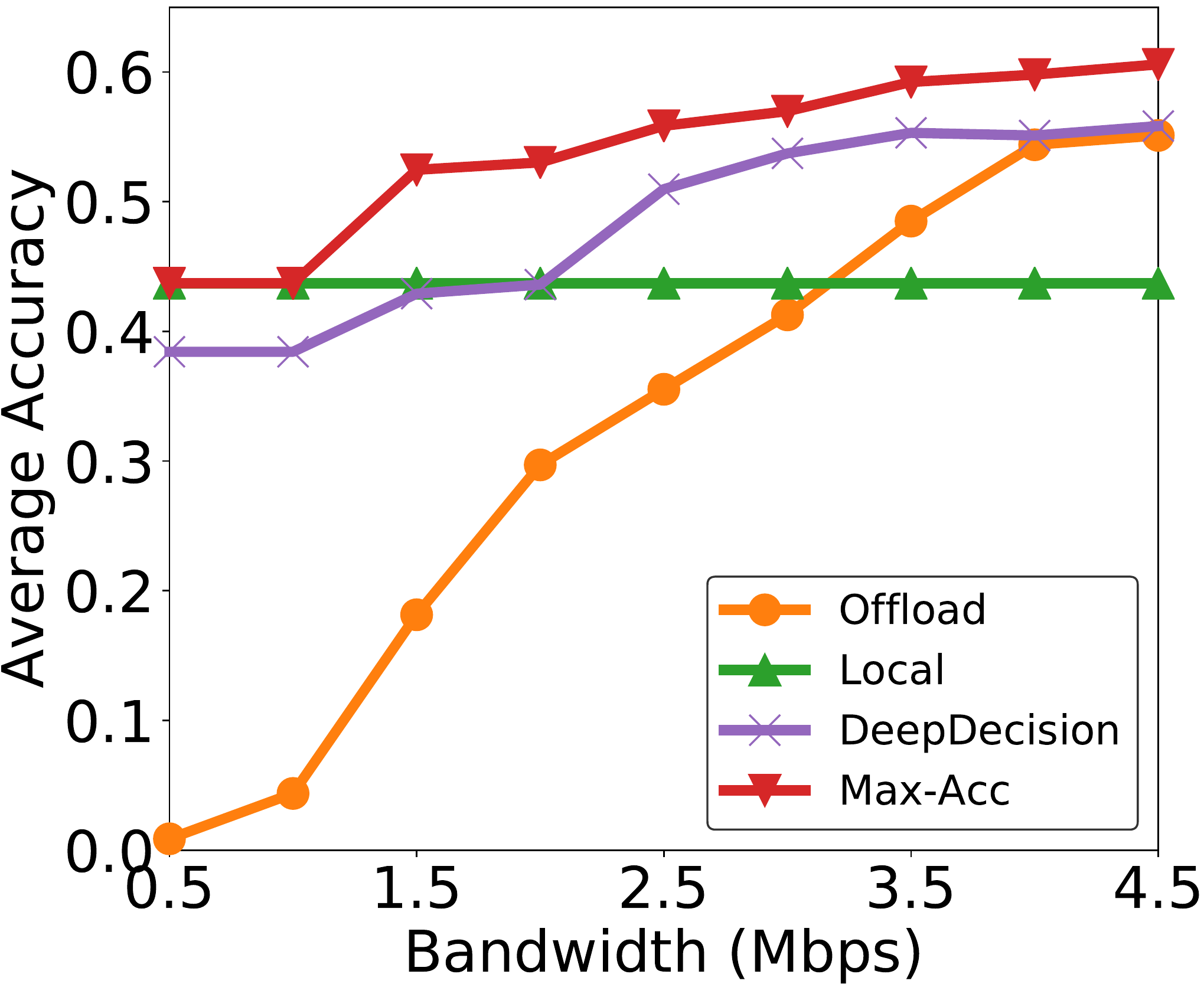}
				\label{fig:max-acc-bandwidth-1}}}
		\subfloat[Frame Rate 50 fps]{{
				\includegraphics[width=0.49\linewidth, valign=t]{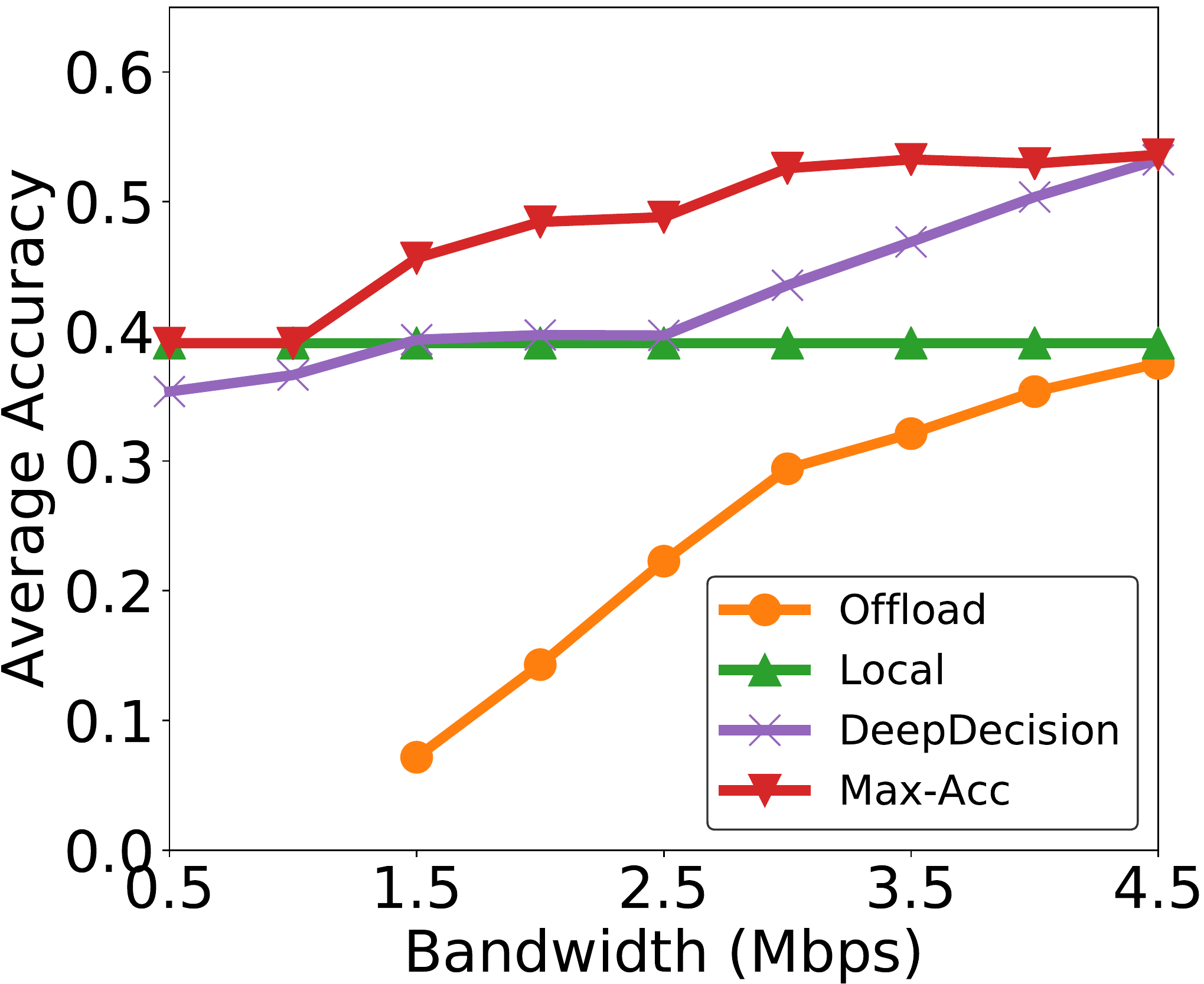}
				\label{fig:max-acc-bandwidth-2}}}
		\caption{The performance of different methods under different network conditions.}
		\label{fig:exp-max-acc-bandwidth}
		\vspace{-0.5cm}
	\end{figure}
	
	In Figure \ref{fig:exp-max-acc-bandwidth}, we compare Max-Accuracy with the Local and Offload method under different network conditions. In the evaluation, we set the frame upload delay to be 100 ms.
	The Local method does not offload any frames, and thus its performance remains the same under different network conditions.
	The Local method can achieve the same accuracy as the Max-Accuracy algorithm when the bandwidth is low, since most of the video frames will be processed locally and the Local method can find an optimal solution.
	Notice that the Local method performs better than DeepDecision when the bandwidth is low.
	The reason is as follows. 
	DeepDecision makes the same schedule decision for frames within a time slot and NPU may not be fully utilized if only SqueezeNet is used.
	In contrast, the Local method achieves higher accuracy by using ResNet to process some of the frames within the time slot.
	In Figure \ref{fig:exp-max-acc-bandwidth}(b), the Offload method is not capable of processing all frames when the bandwidth is lower than 1.5 Mbps.
	When the network bandwidth is low, the Offload method performs poorly since it has to resized video frames into an extremely small size and then reduce the accuracy.
	As shown in Figure \ref{fig:resolutions}, even with an advanced CNN model, the accuracy is till low with these low resolution images. 
	As the network bandwidth increases, the differences among the Max-Accuracy, DeepDecision and Offload become smaller since the mobile device can offload most of the frames in high resolution and achieve better accuracy. 
	
	\begin{figure}
		\centering
		\subfloat[$B = 2$ Mbps]{{
				\includegraphics[width=0.49\linewidth]{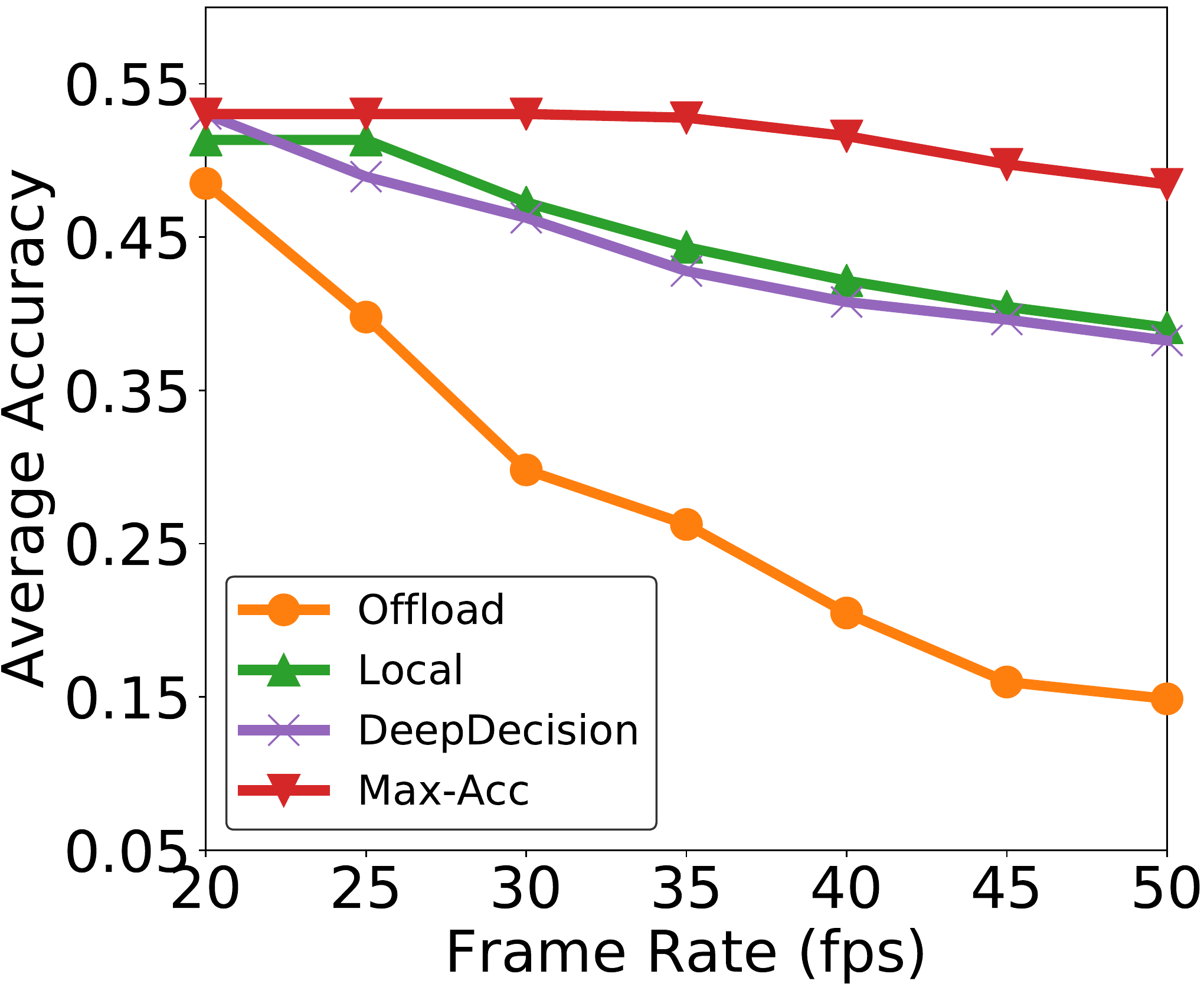}
				\label{fig:max-acc-fps-1}}}
		\subfloat[$B = 3$ Mbps]{{
				\includegraphics[width=0.49\linewidth]{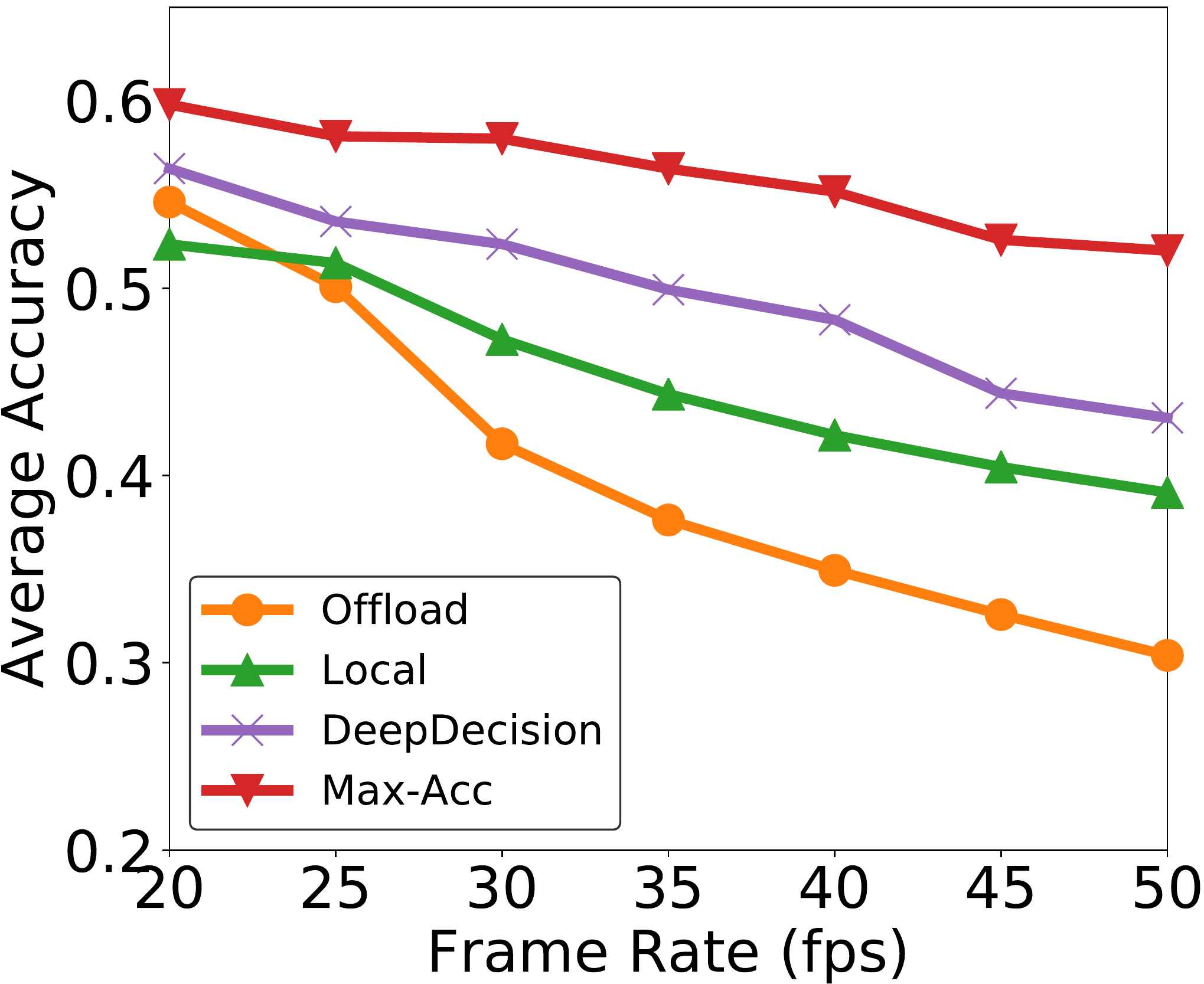}
				\label{fig:max-acc-fps-2}}}
		\caption{The performance of different methods under different frame rate requirements.}
		\label{fig:exp-max-acc-fps}
		\vspace{-0.7cm}
	\end{figure}
	
	In Figure \ref{fig:exp-max-acc-fps}, we evaluate the impact of frame rate for different methods.
	As can be seen from the figures, the performance of all methods drops when the frame rate is high.
	As the frame rate requirement increases, more frames have to be resized to lower resolutions. 
	That is why the  Offload method suffers a 30\% accuracy drop in the experiments.
	In contrast, there is no significant accuracy drop in Max-Accuracy, since they can avoid reducing the solution by processing the video frames on NPU. 
	
	\begin{figure}
		\centering
		\subfloat[Performance of the Optimal]{{
				\includegraphics[width=0.49\linewidth]{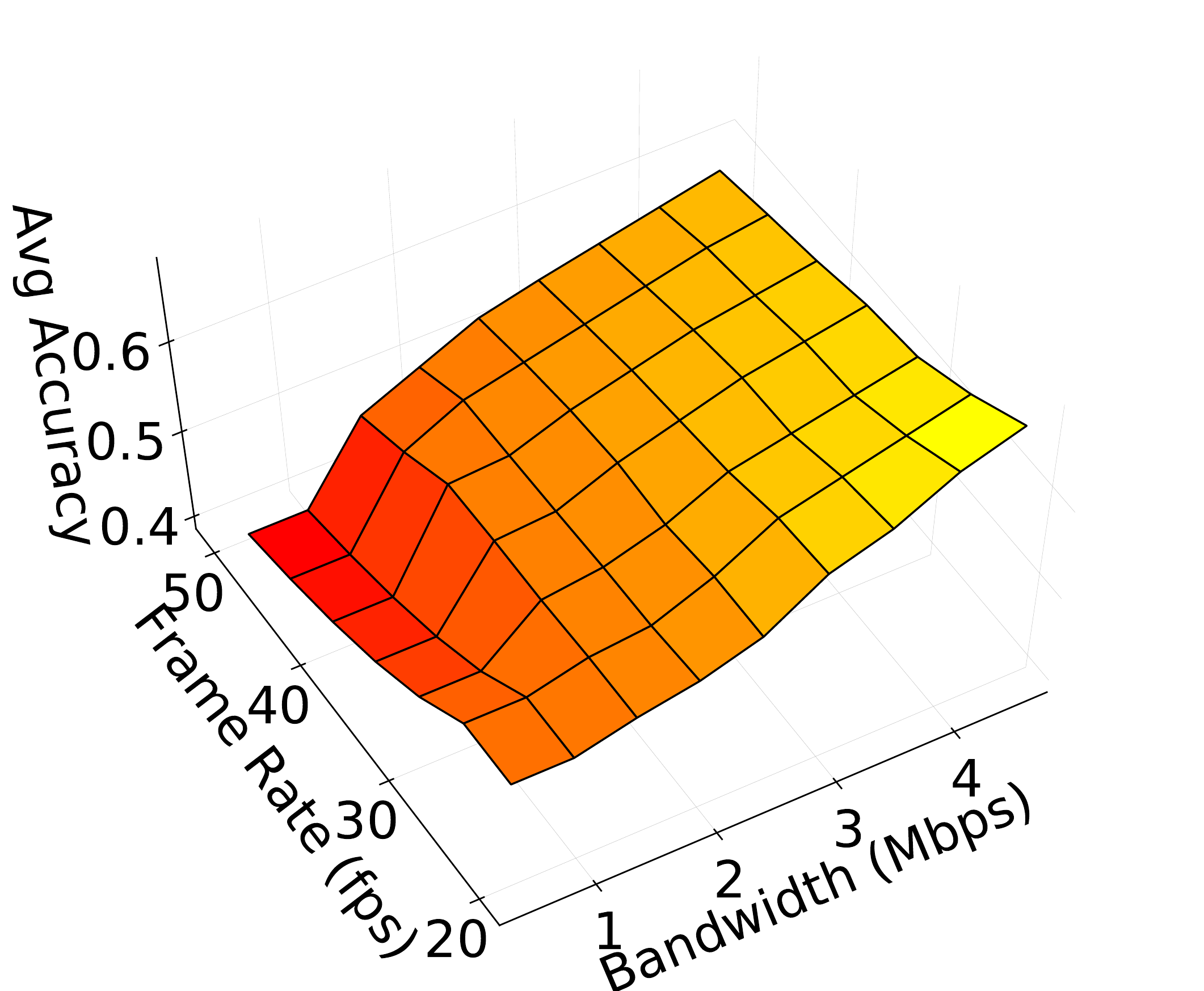}
		}}
		\subfloat[Difference between Max-Accuracy and Optimal]{{
				\includegraphics[width=0.49\linewidth]{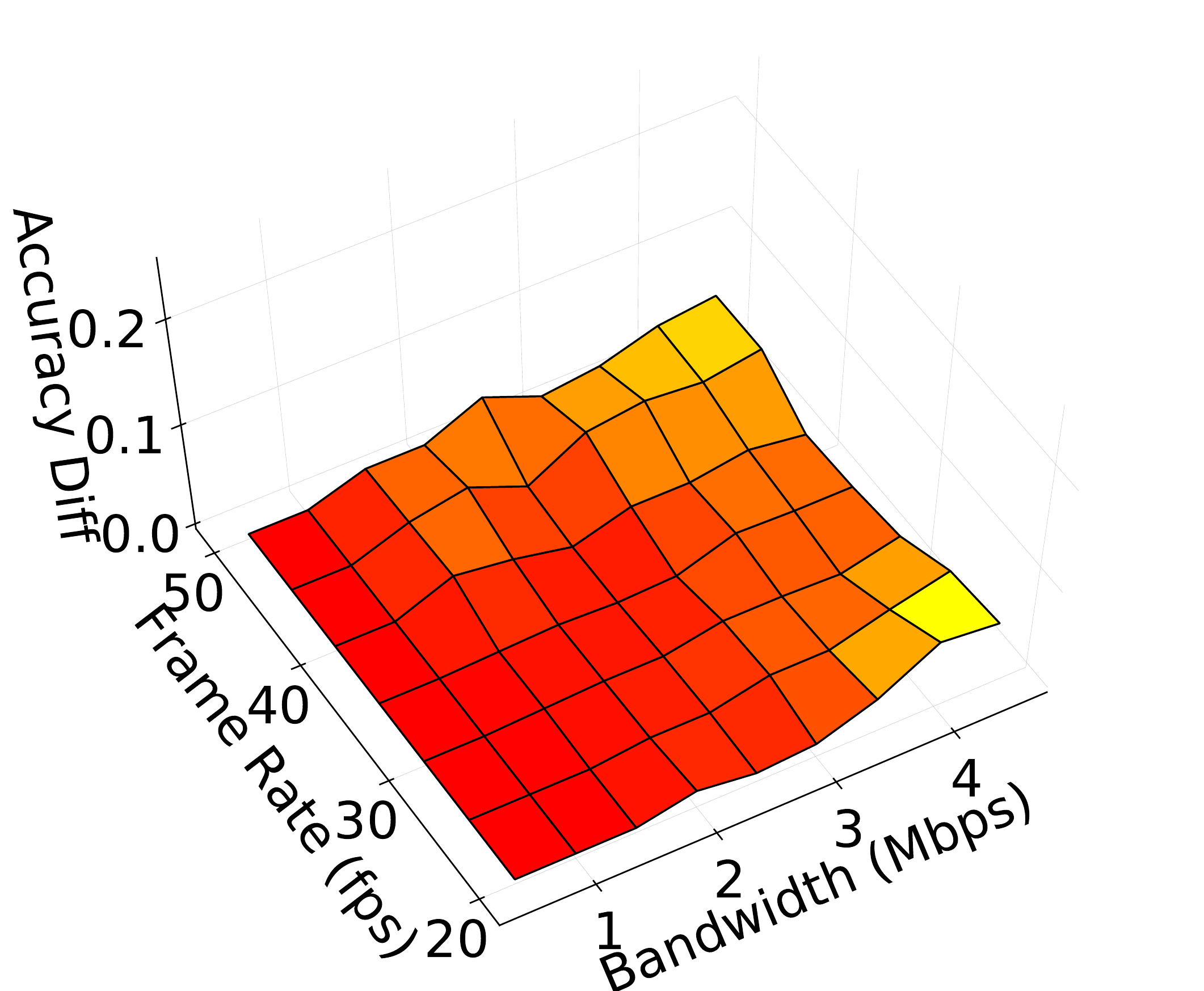}
		}}
		\caption{Comparison between optimal and Max-Accuracy}
		\label{fig:exp-max-acc-3d}
		\vspace{-0.5cm}
	\end{figure}
	
	In Figure \ref{fig:exp-max-acc-3d}, we compare Max-Accuracy with the Optimal method under various frame rates and network conditions. 
	As shown in Figure \ref{fig:exp-max-acc-3d}(a), the accuracy of Optimal increases when the network bandwidth increases, because 
	the mobile device can upload more frames with higher resolution.
	To support a higher frame rate, more frames must be processed within the time constraint and the optimal method has to use fast CNN models with low resolution or low accuracy, resulting in low accuracy. 
	
	In figure \ref{fig:exp-max-acc-3d}(b), we plot the accuracy difference between Optimal and Max-Accuracy.
	The accuracy difference is computed using the accuracy of the Optimal method minus that of Max-Accuracy.
	As can be seen from the figure, the difference is almost 0 in most cases, which indicates that Max-Accuracy is close to Optimal.
	
\begin{figure}
		\centering
		\subfloat[Frame Rate 30 fps]{{
				\includegraphics[width=0.49\linewidth]{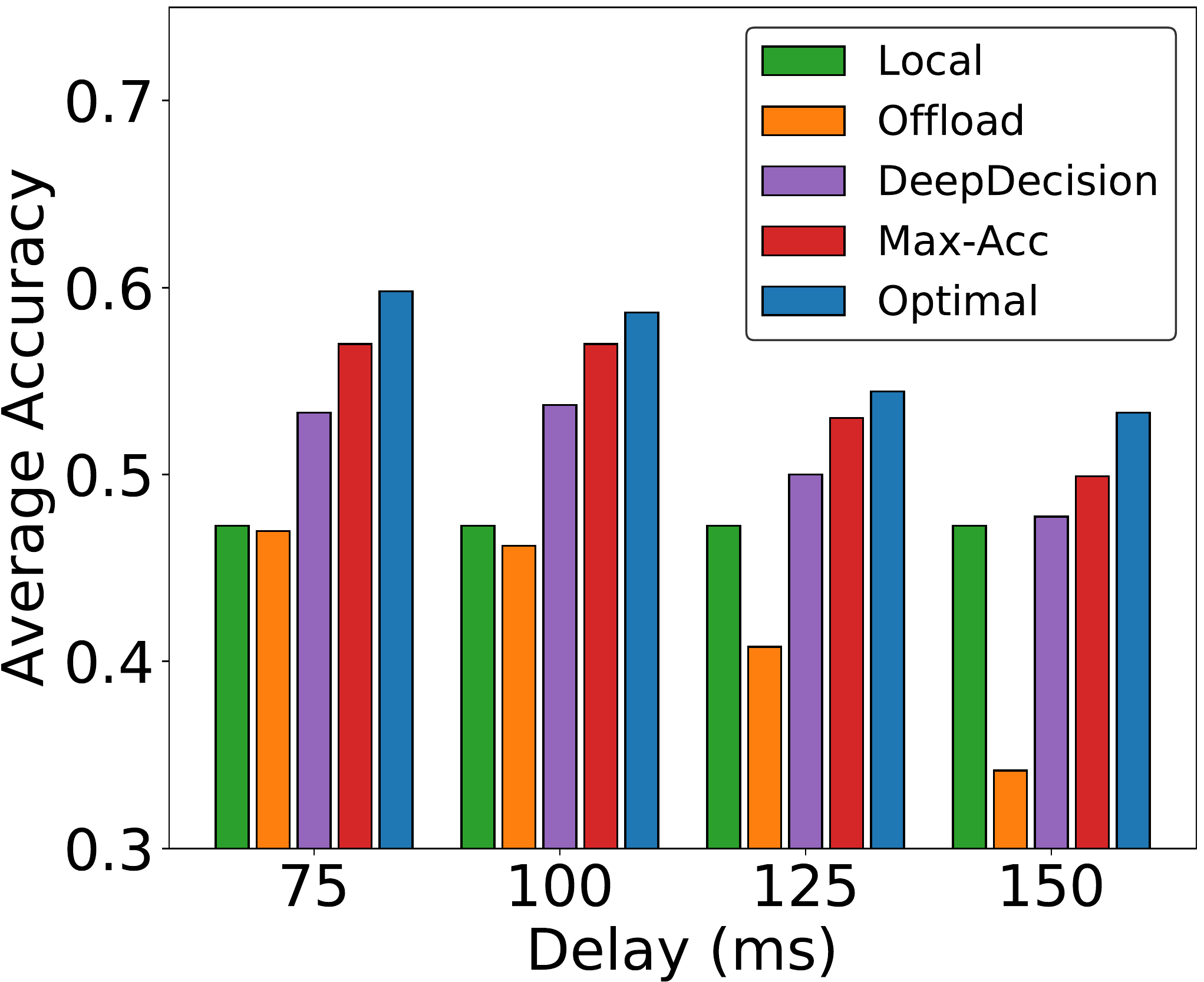}
				\label{fig:max-acc-latency-1}}}
		\subfloat[Frame Rate 50 fps]{{
				\includegraphics[width=0.49\linewidth]{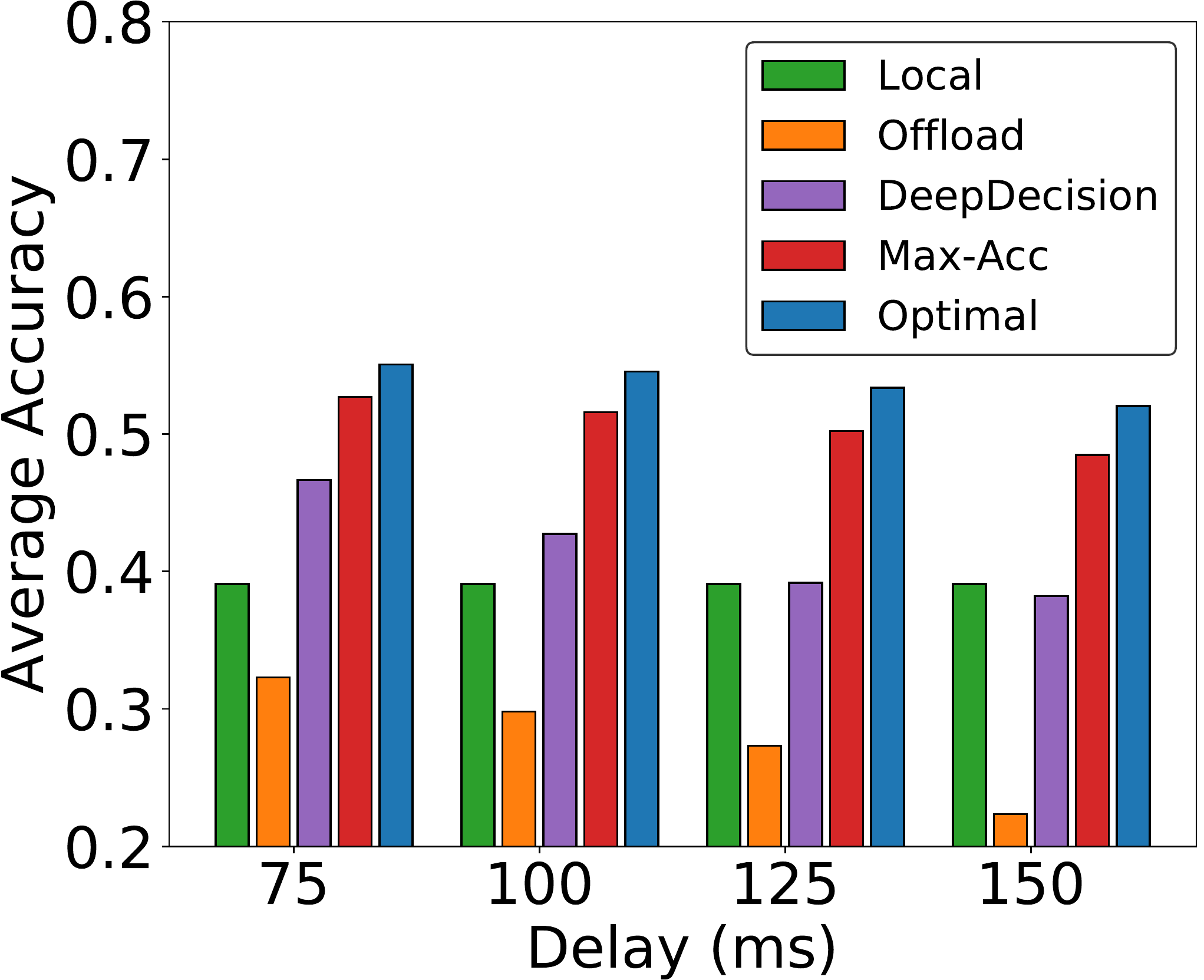}
				\label{fig:max-acc-latency-2}}}
		\caption{The performance of different methods under different frame upload delay.}
		\label{fig:exp-max-acc-latency}
		\vspace{-0.8cm}
	\end{figure}

	In Figure \ref{fig:exp-max-acc-latency}, we evaluate the impact of frame upload delay on accuracy. 
	We set the uplink network bandwidth to be 3 Mbps and set the frame rate to be 30 and 50 fps.
	Since the Local method does not offload any frames, its performance remains the same.
	A longer delay means that less frames can be offloaded to the server for processing, since the result must be returned within the time constraint.
	Therefore, for the Offload, DeepDecision, Optimal and Max-Accuracy algorithms, the performance drops as the upload delay increases.
	Compared to DeepDecision, Optimal and Max-Accuracy, a significant accuracy drop can be observed in the Offload method, when the upload delay becomes larger. This is because DeepDecision, Optimal and Max-Accuracy can schedule frames to be processed locally at this time to deal with the long upload delay. Although the accuracy is also dropped by processing the frames on NPU, the impact is not as significant as that in the Offload method.

	\subsection{The Performance of Max-Utility}
	
	To evaluate the performance of Max-utility, we still compare it to Offload, Local, and Optimal. Since we focus on utility instead of accuracy in this subsection, these algorithms are also modified to maximize utility instead of accuracy.
	
	\begin{figure}
		\centering
		\subfloat[$\alpha=200$]{{
				\includegraphics[width=0.49\linewidth]{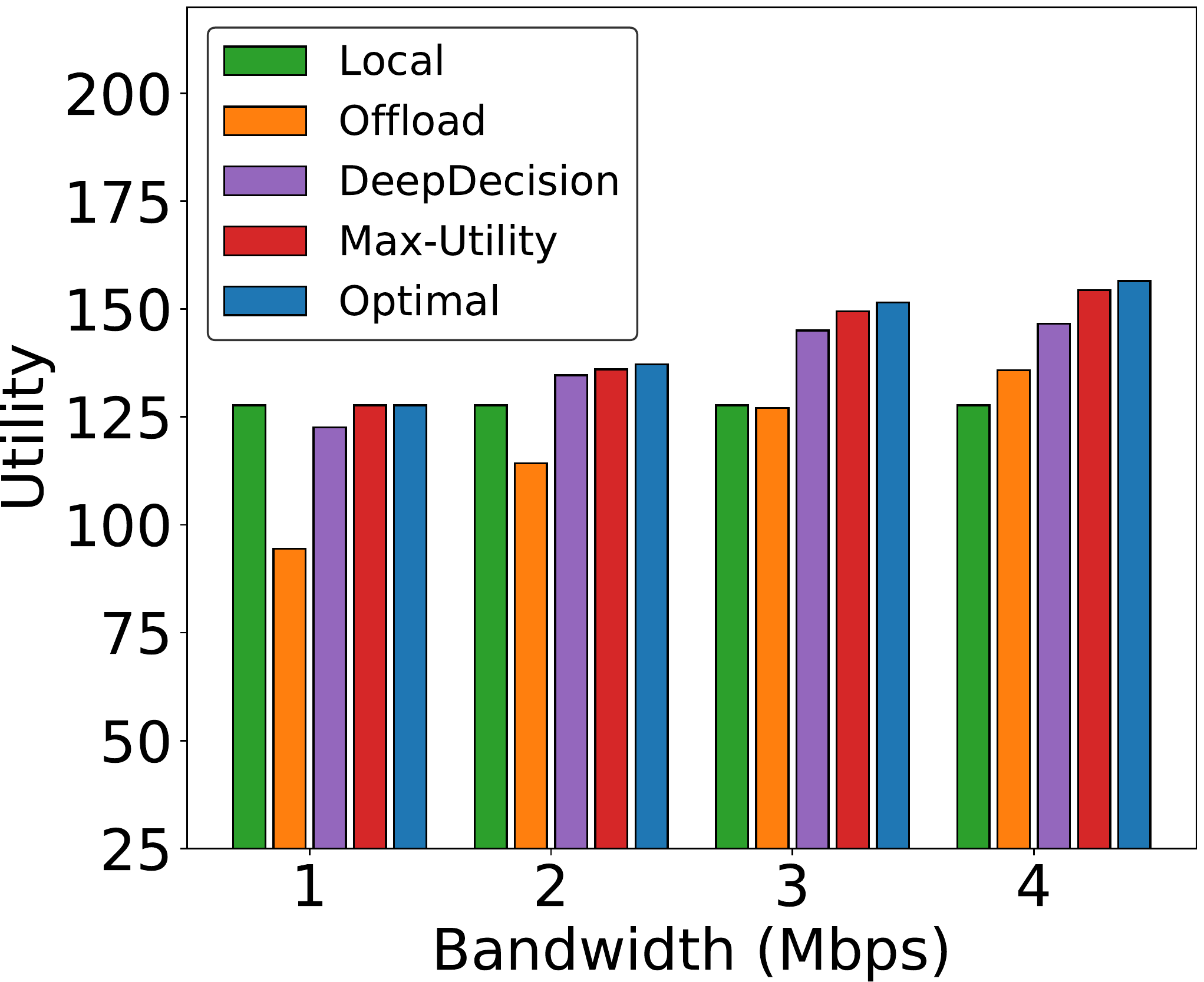}\label{fig:max-utl-bandwidth-1}}}
		\subfloat[$\alpha=50$]{{
				\includegraphics[width=0.49\linewidth]{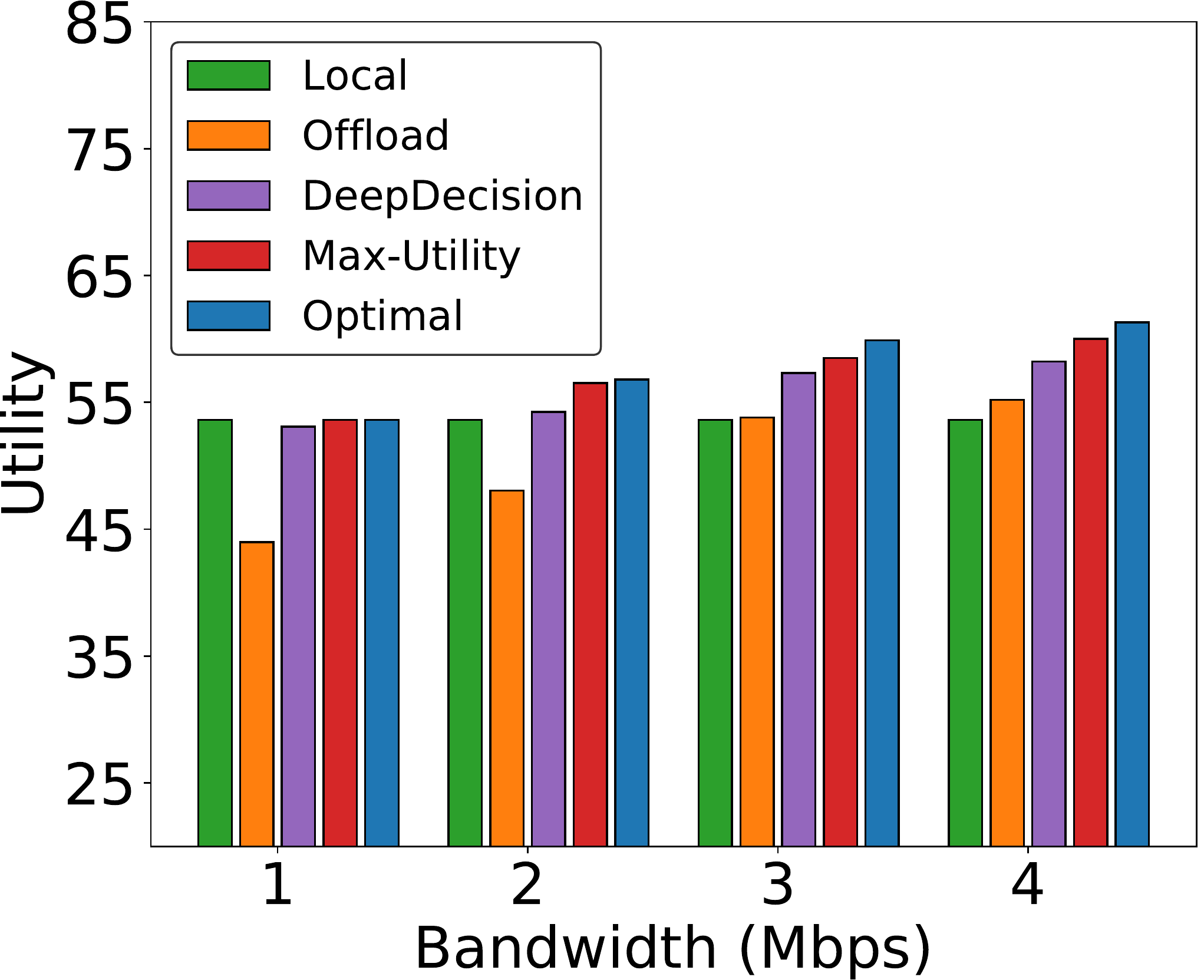}\label{fig:max-utl-bandwidth-2}}}
		\caption{The impact of network bandwidth.}
		\label{fig:exp-max-utl-bandwidth}
		\vspace{-0.5cm}
	\end{figure}
	
	In Figure \ref{fig:exp-max-utl-bandwidth}, we evaluate the impact of network bandwidth for different methods.
	In the comparison, the frame rate and the upload delay are set to be 30 fps and 100 ms respectively, and the tradeoff parameter $\alpha$ is set to be 50 and 200.
	The Local approach cannot offload any video frames to the server, and thus its utility remains the same in different network conditions.
	When the bandwidth is low, the Offload method may have to upload low resolution frames, resulting in low utility, and thus it underperforms the Local method. 
	As the network bandwidth increases, the performance difference among Offload, DeepDecision, Max-Utility, and Optimal becomes smaller, since most frames can be transmitted with higher resolution to achieve better accuracy. 
	
	As shown in the figure, when the network bandwidth decreases, the performance of Offload, DeepDecision, Max-Utility and Optical all drops. 
	However, the performance of DeepDecision, Max-Utility and Optimal drops much slower than Offload. 
	The reason is as follows. 
	When $\alpha$ is large, as shown Figure  \ref{fig:exp-max-utl-bandwidth}(a), the accuracy has more weight in calculating 
	the utility.  Max-Utility achieves high accuracy and then high utility by offloading when network bandwidth is high and by local execution when the network bandwidth is low. 
	When $\alpha$ is small, as shown Figure  \ref{fig:exp-max-utl-bandwidth}(b), the processing time (frame rate) has more weight in calculating the utility.  Max-Utility supports high frame rate and then achieves high utility by offloading when network bandwidth is high and by local execution when the network bandwidth is low.

	\begin{figure}
		\centering
		\subfloat[$\alpha=200$]{{
				\includegraphics[width=0.49\linewidth]{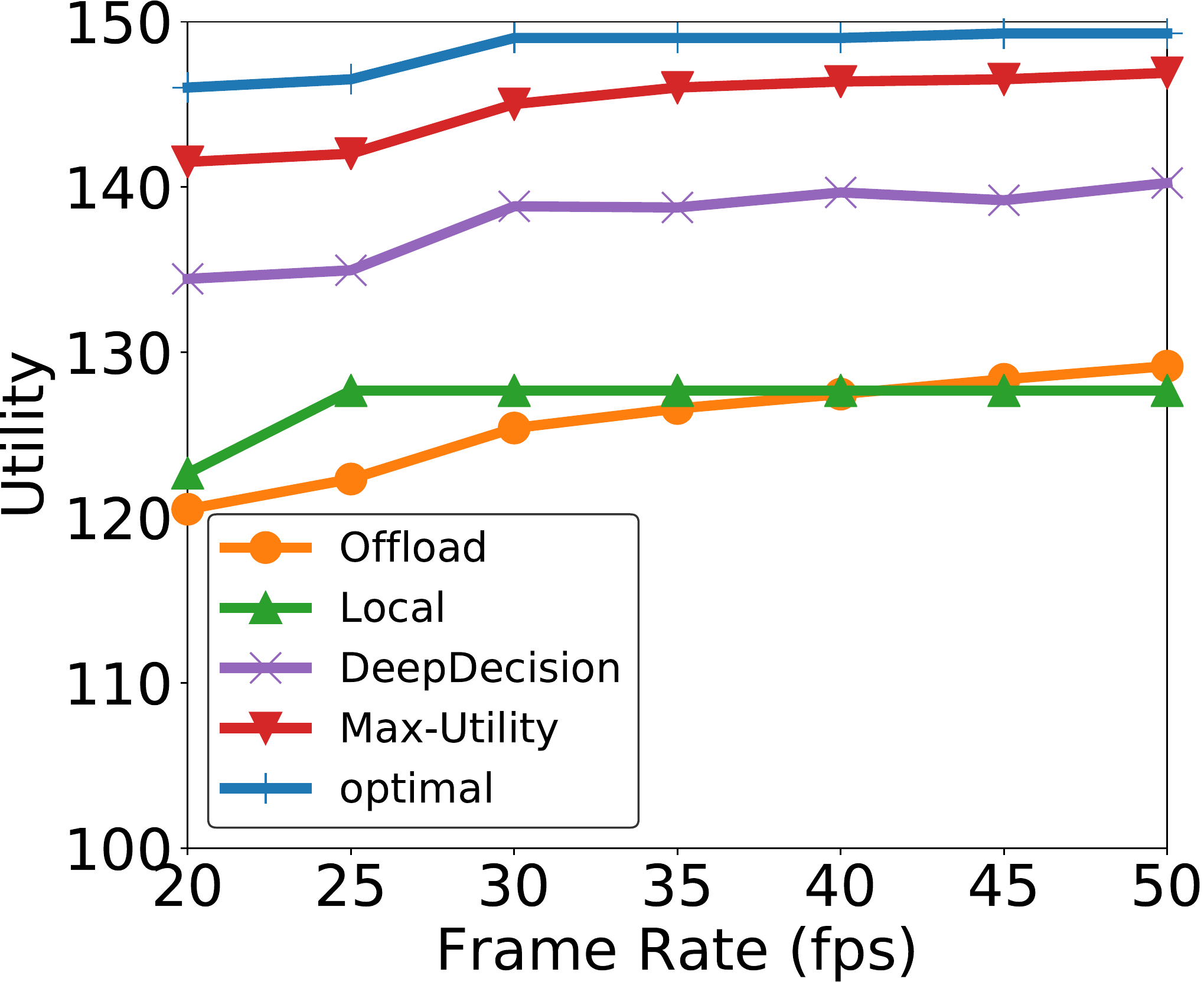}}}
		\subfloat[$\alpha=50$]{{
				\includegraphics[width=0.49\linewidth]{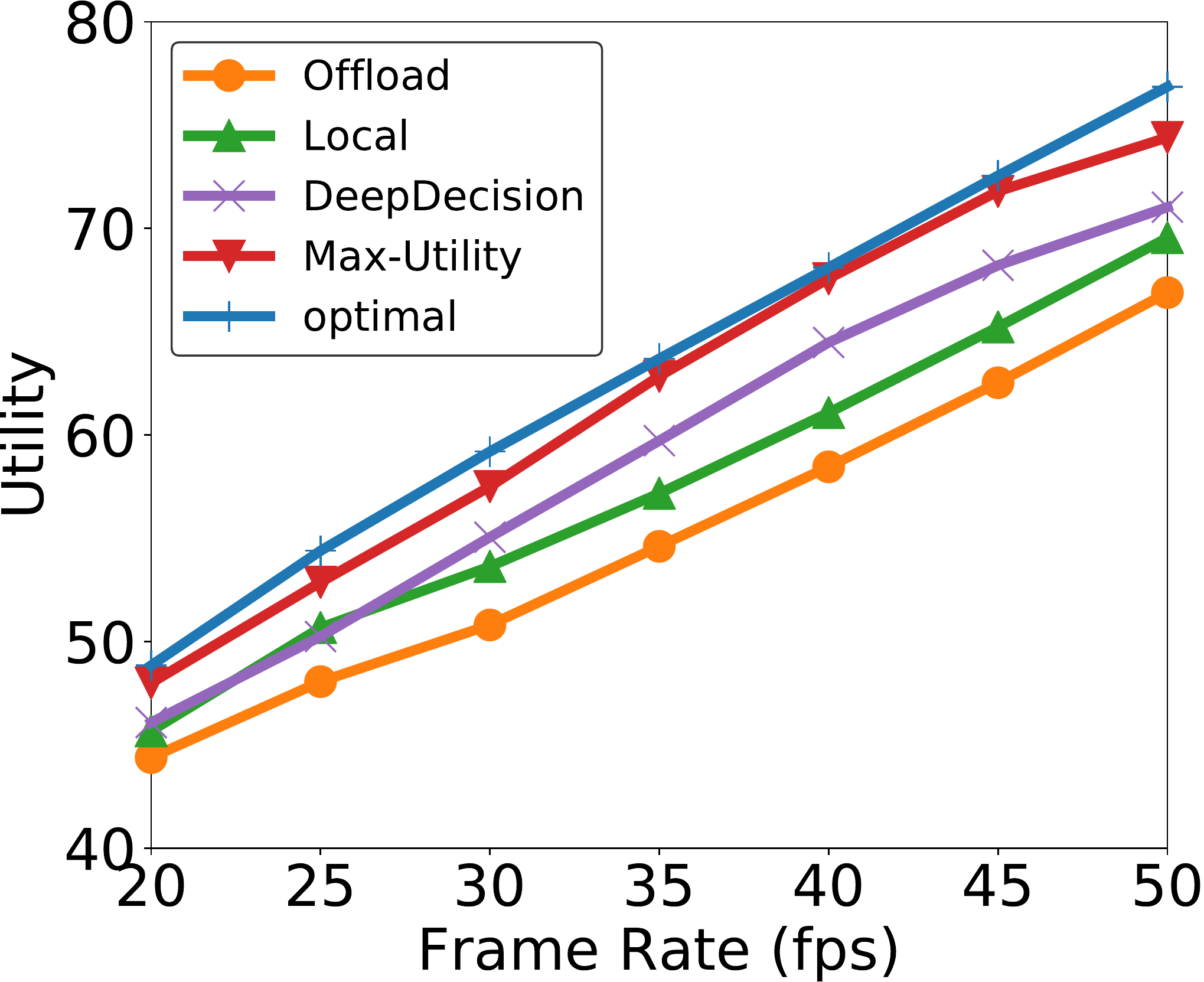}}}
		\caption{The effects of frame rates.}
		\label{fig:exp-max-utl-fps}
 		\vspace{-0.27cm}
	\end{figure}
	
	Figure \ref{fig:exp-max-utl-fps} shows the impacts of frame rate for different methods.
	In the evaluation, we set the network bandwidth to be 2.5 Mbps and the upload frame delay to be 100 ms.
	As shown in the figure, Max-Utility outperforms Offload, Local and DeepDecision methods. 
	When $\alpha$ is small, as shown Figure  \ref{fig:exp-max-utl-fps}(b), the processing time (frame rate) has more weight in calculating the utility, and thus the utility of all methods increases when the frame rate increases.  
	When $\alpha$ is large, as shown Figure  \ref{fig:exp-max-utl-fps}(a), the accuracy has more weight in calculating the utility, and thus the utility of all methods does not increases too much when the frame rate increases.

	\begin{figure}
		\centering
		\subfloat[$\alpha=200$]{{
				\includegraphics[width=0.49\linewidth]{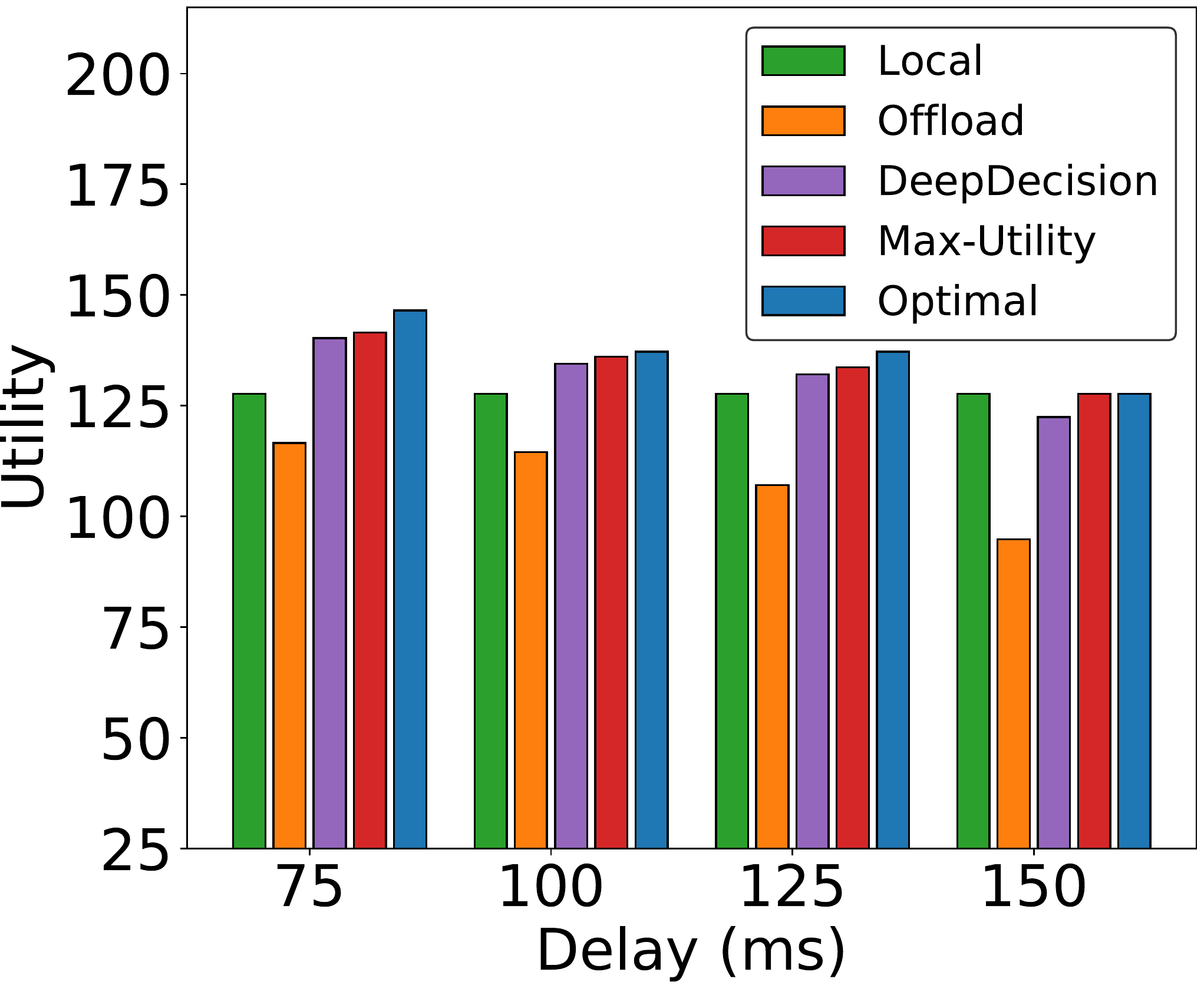}}}
		\subfloat[$\alpha=50$]{{
				\includegraphics[width=0.49\linewidth]{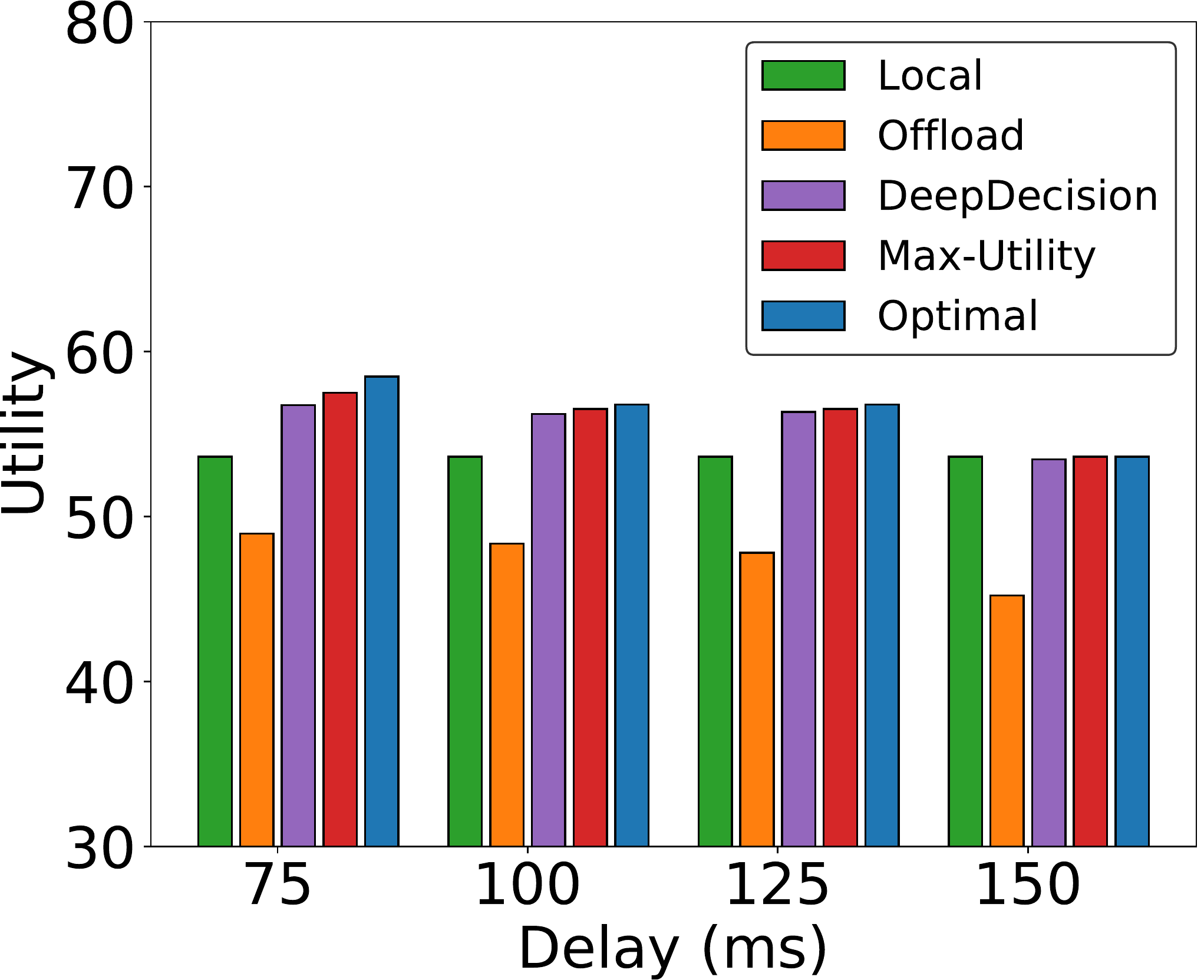}}}
		\caption{The effects of upload delay}
		\label{fig:exp-max-utl-latency}
		\vspace{-0.5cm}
	\end{figure}
	
	Figure \ref{fig:exp-max-utl-latency} shows the impact of upload delay for different methods.
	We set the frame rate to be 30 fps and the network bandwidth to be 2 Mbps.
	Since the Local method does not offload any video frames to the server, its performance remains the same.
	As the upload delay increases, less video frames can be offloaded to the server due to time constraints, and hence degrading the performance of the 
	Offload, DeepDecision, Max-Utility, and Optimal algorithms.

\section{Conclusions} \label{sec:conclusion}

In this paper, we proposed a framework called FastVA, which supports deep learning video analytics through edge processing and Neural 
Processing Unit (NPU) in mobile.  
We are the first to study the benefits and limitations of using NPU to run CNN models to better understand the characteristics of NPU in mobile.
Based on the accuracy and processing time requirement of the mobile application, we studied two problems: 
\textit{Max-Accuracy} where the goal is to maximize the accuracy under some time constraints, and 
\textit{Max-Utility} where the goal is to maximize the utility which is a weighted function of processing time and accuracy. 
To solve these two problems, we have to determine when to offload the computation and when to use NPU. The solution depends on the network condition, the special characteristics of NPU, and the optimization goal. 
We formulated them as integer programming problems and proposed heuristics based solutions. 
We have implemented FastVA on smartphones and demonstrated its effectiveness through extensive evaluations.

\bibliographystyle{IEEEtran}
\bibliography{sample-base}

\end{document}